\documentclass[preprint,12pt,eqsecnum]{aastex}           
\usepackage{natbib,graphics}

\newcommand{\be}{\begin{equation}}
\newcommand{\ee}{\end{equation}}

\newcommand{\bea}{\begin{eqnarray}}
\newcommand{\eea}{\end{eqnarray}}
\newcommand{\nd}{\nodata}

\slugcomment{Submitted to The Astrophysical Journal on April 7, 2000}
\shorttitle{Relativistic Diskoseismology. II: C--modes}
\shortauthors{Silbergleit, Wagoner \& Ortega-Rodr\'{\i}guez}

\begin{document}
\title{Relativistic Diskoseismology. II. Analytical Results for C--modes} 

\author{Alexander S. Silbergleit\altaffilmark{1}}
\affil{Gravity Probe B, Stanford University, Stanford, CA 94305--4085}
\author{Robert V. Wagoner\altaffilmark{2}} 
\affil{Department of Physics and Center for Space Science and Astrophysics \\ 
Stanford University, Stanford, CA 94305--4060 \\ and Institute for Theoretical Physics, U.C. Santa Barbara}
\and\author{Manuel Ortega-Rodr\'{\i}guez\altaffilmark{3}}
\affil{Department of Applied Physics and Gravity Probe B, \\
Stanford University, Stanford, CA 94305--4090}

\altaffiltext{1}{gleit@relgyro.stanford.edu} 
\altaffiltext{2}{wagoner@stanford.edu} 
\altaffiltext{3}{manuel@leland.stanford.edu}

\begin{abstract}

We first briefly review how we investigate the modes of oscillation trapped within the inner region of accretion disks by the strong-field gravitational properties of a black hole (or a compact, weakly-magnetized neutron star). Then we focus on the `corrugation'(c)--modes, nearly incompressible perturbations of the inner disk. The fundamental c--modes have eigenfrequencies (ordered by radial mode number) which correspond to the Lense--Thirring frequency, evaluated at the outer trapping radius of the mode, in the slow rotation limit. This trapping radius is a decreasing function of the black hole angular momentum, so a significant portion of the disk is modulated only for slowly rotating black holes. The eigenfrequencies are thus strongly increasing functions of black hole angular momentum. The dependence of the eigenfrequencies on the speed of sound within (or the luminosity) within the disk is very weak, except for slowly rotating black holes.

\end{abstract}


\section{Introduction}

For twenty years [beginning with \citet{kf}], it has been known that general relativity can trap normal modes of oscillation near the inner edge of accretion disks around black holes. The strong gravitational fields that are required can also be produced by neutron stars that are sufficiently compact (with a soft equation of state) and weakly magnetized to produce a gap between the surface of the star and the innermost stable orbit of the accretion disk. Although we shall not explicitly consider such neutron stars here, the results obtained will also apply to them to first order in the dimensionless angular momentum parameter $a=cJ/GM^2$, since their exterior metric is identical to that of a black hole to that order. 

These modes of oscillation provide a potentially powerful probe of both strong gravitational fields and the physics of accretion disks, since they do not exist in Newtonian gravity. In addition, their frequencies depend upon the angular momentum as well as the mass of the black hole. The fractional frequency spread of each mode depends upon the elusive viscosity parameter of the accretion disk.

The subject of `relativistic diskoseismology' has recently been reviewed by \citet{w}. 
In this paper we shall focus on the `corrugation'(c) modes, previously studied (less generally) by \citet{k90,k93}, \citet{i94,i96}, and \citet{p}. We shall briefly compare these modes with the `gravity'(g) modes and the pressure (p) modes.

\section{Basic Assumptions and Equations}

\subsection{Structure of the unperturbed accretion disk}

We take $c=1$, and express all distances in units of $GM/c^2$ and all frequencies in units of $c^3/GM$ (where $M$ is the mass of the central body) unless otherwise indicated. We employ the Kerr metric to study a thin accretion disk, neglecting its self-gravity. The stationary ($\partial/\partial t=0$), symmetric about the midplane $z=0$, and axially symmetric ($\partial/\partial\varphi$=0) equilibrium disk is taken to be described by the standard relativistic thin accretion disk model \citep{nt,pt}. The velocity components $v^r=v^z=0$, and the disk semi-thickness $h(r)\sim c_s/\Omega\ll r$, where $c_s(r,z)$ is the speed of sound. The key frequencies, associated with free-particle orbits in the disk, are 
\bea
\Omega(r) & = & (r^{3/2}+a)^{-1}\; , \nonumber \\
\Omega_\perp(r) & = & \Omega(r)\left(1-4a/r^{3/2}+3a^2/r^2\right)^{1/2}\; , \nonumber \\
\kappa(r) & = & \Omega(r)\left(1-6/r+8a/r^{3/2}-3a^2/r^2\right)^{1/2}\; ;\label{eq:1}
\eea
the rotational, vertical epicyclic, and radial epicyclic angular velocities, respectively. The angular momentum parameter $a$ is less than unity in absolute value. 

The inner edge of the disk is at approximately the radius of the last stable free-particle circular orbit $r=r_i(a)$, where the epicyclic frequency $\kappa(r_i)=0$. This radius $r_i(a)$ is a decreasing function of $a$, from $r_i(-1)=9$ through $r_i(0)=6$ to $r_i(1)=1$. So all the relations we use are for $r>r_i$, where $\kappa(r)>0$. Note, in particular, that
\be
\Omega(r)>\Omega_\perp(r)>\kappa(r) \; ,\quad a>0 \; ;\qquad \Omega_\perp(r)>\Omega(r)>\kappa(r) \; ,\quad a<0 \; .
\label{eq:2}
\ee
For $a=0$, i.~e., a non-rotating disk, $\Omega(r)=\Omega_\perp(r)>\kappa(r)$.

To simplify the analysis, we here consider mostly barotropic disks [$p=p(\rho)$, vanishing buoyancy frequency; a generalization to a small non-zero buoyancy is described briefly in section 4.3]. In this case hydrostatic equilibrium provides the vertical density and pressure profiles
\be
\rho=\rho_0(r)(1-y^2)^{1/(\Gamma-1)}\; ,\quad p=p_0(r)(1-y^2)^{\Gamma/(\Gamma-1)} \qquad (\Gamma >1) \; . \label{eq:3}
\ee
The disk surfaces are at $y=\pm1$, with $y$ related to the vertical coordinate $z$ by
\[ y={z\over h(r)}\,\sqrt{\Gamma-1\over2\Gamma} \; . \]
The adiabatic index $\Gamma=4/3$ within any radiation pressure dominated region of the disk, and $\Gamma=5/3$ within any gas pressure dominated region. More information on the unperturbed disk can be found in sections 4.3 and 5.1.

\subsection{Equations for the disk perturbations}

To investigate the eigenmodes of the disk oscillations, we apply the general relativistic formalism that \citet{il} developed for perturbations of purely rotating perfect fluids. Neglecting self-gravity, it allows one to express Eulerian perturbations of all physical quantities through a single function $\delta V\propto\delta p/\rho$ which satisfies a second-order partial differential equation. Due to the stationary and axisymmetric background, the angular and time dependences are factorized out as $\delta V = V(r,z)\exp[i(m\phi + \sigma t)]$, where $\sigma$ is the eigenfrequency. Then the assumption of strong variation of modes in the radial direction (characteristic radial wavelength $\lambda_r \ll r$) ensures the WKB separability of variables in the partial differential equation for the functional amplitude $V(r,z) = V_r(r)V_y(r,y)$. The function $V_y$ varies slowly with $r$. The resulting ordinary differential equations for the vertical ($V_y$) and radial ($V_r$) eigenfunctions are [see \citet{nw92} and \citet{per} for details]:  
\bea
(1-y^2)\frac{d^2V_y}{d y^2} - \frac{2y}{\Gamma -1}\frac{d V_y}{d y} + \frac{2\omega_*^2}{\Gamma -1}\left[1 + 
\frac{\Psi-\omega_*^2}{\omega_*^2}\left(1-y^2\right)\right]V_y = 0 \; ,
\label{eq:4} \\
\frac{d^2 V_r}{dr^2} - \frac{1}{(\omega^2-\kappa^2)} \left[\frac{d}{dr}
(\omega^2-\kappa^2)\right]\frac{dV_r}{dr} +\alpha^2(\omega^2-\kappa^2)\left(1 -
\frac{\Psi}{\omega_*^2}\right)V_r = 0 \; . \label{eq:5}
\eea 
The coefficient $\alpha(r)$ coincides with $1/c_s(r,0)$ up to a relativistic factor of order unity which varies slowly with radius.

Together with the appropriate homogeneous boundary conditions [discussed by \citet{per}], these equations generate the vertical and radial eigenvalue problems, respectively.  The radial  conditions depend on the type of mode and its capture zone, as discussed in the following sections. The coefficient $\alpha$, the vertical eigenfunction $V_y$ and  eigenvalue (separation function) $\Psi$ vary slowly with radius, as does the dimensionless ratio of the corotation frequency $\omega(r)$ and $\Omega_\perp(r)$: 
\be
\omega_*(r)\equiv\omega(r)/\Omega_\perp(r) \; , \qquad\omega(r)\equiv\sigma+m\Omega(r) \; . \label{eq:6}
\ee
(Note that $\Psi$ and $\omega_*$ depend also on the eigenfrequency $\sigma$.)

Our goal is to determine the vertical eigenvalues $\Psi(r)$ and the corresponding spectrum of eigenfrequencies $\sigma$ from the two eigenvalue problems for equations (\ref{eq:4}) and (\ref{eq:5}).

\section{Classification of Modes}

One can see from the radial equation (\ref{eq:5}) that a mode oscillates in a domain of $r>r_i$ where the quantity $(\omega^2-\kappa^2)(1 -\Psi/\omega_*^2)$ is positive. The first factor here (for $\sigma$ in the allowed range) changes sign twice at points $r=r_{\pm}(m,a,\sigma),\,\,r_i<r_-\leq r_+ $, determined by equations (1) and (\ref{eq:6}). This factor is negative between $r_-$ and $r_+$ and positive otherwise. In Figure 1 radii $r_{\pm}$ are plotted as functions of $\sigma$ for a few values of $m$ and $a$. Thus the classification of modes is based on the behavior of the second factor, containing the eigenvalue. 

There are essentially three possibilities, corresponding to the following types of modes: 

\noindent$\bullet$ {\it g--modes}: $\Psi>$(typically\ $\gg )\,\omega_*^2$; capture zone $r_-<r<r_+$. 

\noindent$\bullet$ {\it p--modes}: $\Psi<$(typically\ $\ll )\,\omega_*^2$; capture zone $r_i<r<r_-$ or $r>r_+$.

\noindent$\bullet$ {\it c--modes}: $\Psi = \omega_*^2$, at least at one value of the radius. 

\noindent The capture, or trapping, zone is where the mode oscillations occur.

The g (inertial-gravity)--modes are centered on the radius $r_m$ where $r_-=r_+$, corresponding to the maximum eigenfrequency $|\sigma|$. The corotating eigenfrequencies $|\omega|$ corresponding to the lowest mode numbers are close to $\kappa(r_m)$. The p (inertial-pressure)--modes involve the uncertain physics at either the inner or outer radius of the disk.

The c(corrugation)--modes which have so far been investigated are of a special sort; namely, $\Psi-\omega_*^2\approx 0$ throughout their whole capture domain. Only such c-modes are studied in this paper. They are typically non-radial ($m=\pm1$) vertically incompressible waves near the inner edge of the corotating disk that precess around the angular momentum of the black hole. Their fundamental frequency is shown below to coincide with the Lense-Thirring \citep{lt} frequency (produced by the dragging of inertial frames generated by the angular momentum of the black hole) at their capture zone boundary $r_c(m,a)$ in the appropriate slow-rotation limit. This interesting fact is one reason we are concentrating on the c--modes in this paper. The other reason is that g-- and p--modes have been rather extensively studied by \citet{per} and \citet{nw91}, respectively [and in earlier papers cited therein and in a brief summary of results by \citet{w}]. 

\section{Theory of C--modes and the Lense--Thirring Frequency}

\subsection{Vertical eigenvalue problem: a selection rule for the axial and vertical mode numbers}

According to the definition of the c--modes we are interested in, we set
\be
\Psi(r)/\omega_*^2(r)=1-\chi(r)\; , \qquad |\chi(r)|\ll 1 \; ,
\label{eq:7}
\ee
and write the vertical equation (3) as
\be
(1-y^2)\frac{d^2V_y}{dy^2} - \frac{2y}{\Gamma -1}\frac{dV_y}{ dy} + \frac{2\omega_*^2}{\Gamma -1}\left[1-\chi(1-y^2)\right]V_y = 0 \; , 
\label{eq:8}
\ee
with the boundary condition $|V_y(\pm1)|<\infty$. The last term containing the slowly varying function $\chi(r)$ is considered a small perturbation. We then immediately see that the unperturbed equation coincides with that for the Gegenbauer polynomials \citep{bat} $C^\lambda_j(y),\,\,\lambda=(3-\Gamma)/2(\Gamma-1)>-1/2,\,\,j=0,1,2,\dots$; which are its only solutions regular at $y=\pm1$. Therefore the coefficient of $V_y$ must coincide with the corresponding eigenvalues: $2\omega_*^2/(\Gamma -1)=j(j+2\lambda)$. However, the left hand side of this equality is a (slowly varying) function of $r$, so it can be at best valid only `to the main order'. To clarify the meaning of that, we make use of the definition (\ref{eq:6}) of $\omega_*$ and the fact that the difference between $\Omega$ and $\Omega_\perp$ is typically very small, except for values of $a$ close to $1$ and simultaneously $r$ close to $r_i\approx1$. So (anticipating that $|\sigma|\ll \Omega_\perp$ in the same regime) we introduce a small parameter $\epsilon$ by setting
\be
\omega_*=\frac {\sigma +m\Omega}{\Omega_\perp}\equiv m(1+\epsilon),\quad
\epsilon=\epsilon(r,\sigma,a)=\frac {\sigma +m\left(\Omega-\Omega_\perp\right)}{m\Omega_\perp},\quad
 |\epsilon(r)|\ll 1,
\label{eq:9}
\ee
and expand the eigenvalue and eigenfunction in it:
\be
\chi = \epsilon\chi_1 + {\rm O}(\epsilon^2)\; ,\qquad 
V_y = V^0 + \epsilon V^1 + {\rm O}(\epsilon^2) \; .
\label{eq:10}
\ee
To zero order we therefore have 
\be
V^0(y)\equiv V^0_j(y)=C^\lambda_j(y)
\label{eq:11}
\ee
and $2m^2/(\Gamma -1)=j(j+2\lambda)$, or
\be
2m^2=j^2(\Gamma -1)+j(3-\Gamma)\; ,\qquad m=0,\pm1,\pm2,\dots;\quad j=0,1,2,\dots\; .
\label{eq:12}
\ee
Note that $j$ is the number of vertical eigenfunction nodes through the disk.

Equation (\ref{eq:12}) is the criterion for existence of the studied c-modes. It is also a {\it selection rule} for the angular mode numbers $m$ and $j$ because, depending on the value of $\Gamma$, the modes may not exist for {\it any} $m$ and $j$. On the other hand, given a rational $\Gamma$, there might exist many integer values of $m^2$ and $j$ satisfying equation (\ref{eq:12}). (Using some results from number theory~[\cite{hua}], it is possible to show that equation (\ref{eq:12}) has a {\it finite} set of solutions if and only if $2pq=k^2$ for some integer $k$, with coprime $p,q$ from $\Gamma=1+p/q$. For instance, we have an infinite sequence of c--modes for both $\Gamma=4/3$ and $\Gamma=5/3$, but only a finite number of them if, say, $\Gamma=3/2$.) However, the only solutions {\it independent} of $\Gamma$ are $j=m=0$ (which does not provide radial mode trapping and must thus be discarded) and
\be
j=1=m^2 \; , 
\label{eq:13}
\ee
which specifies the fundamental c-mode. Its vertical eigenfunction to the main order is linear in the vertical coordinate, $V^0_1\propto y$. Some other possible modes are $m^2=4,\; j=3$;  $m^2=25,\; j=10$ for $\Gamma=4/3$; and $m^2=16,\; j=6$ for $\Gamma=5/3$. To be specific, from now on we assume $m>0,\,a>0$, unless indicated otherwise.

When equation (\ref{eq:12}) is satisfied, the first correction $\chi_1$ to the vertical eigenvalue is found in a standard way for the discrete spectrum perturbation, from the solvability condition of the problem for $V^1$:
\be
\chi_1=\chi_1(m,j,\Gamma) =\frac{2\big\langle V^0,V^0\big\rangle}{\big\langle\left(1-y^2\right) V^0,V^0\big\rangle}>0,
\label{eq:14}
\ee
where $\langle\cdot,\cdot\rangle$ is the proper scalar product with the Gegenbauer polynomial weight $\left(1-y^2\right)^{\lambda-1/2}$. From equations (\ref{eq:7}) and (\ref{eq:10}), the eigenvalue is then obtained as
\be
\Psi(r)/\omega_*^2(r)=1 - \epsilon\chi_1 + {\cal O}(\epsilon^2) \; . 
\label{eq:15}
\ee
The integrals in equation (\ref{eq:14}) can be expressed in terms of the Euler gamma function in the general case and are elementary to calculate for $j=1$, giving
\be
\chi_1=3\Gamma-1 \qquad (m^2=j=1) \; .
\label{eq:16}
\ee
The vertical eigenvalue and eigenfunction for all c-modes are completely determined to the needed order by equations (\ref{eq:10}), (\ref{eq:11}), (\ref{eq:12}), (\ref{eq:14}), and (\ref{eq:15}).

\subsection{Radial behavior of the vertical eigenvalue: the Lense-Thirring frequency at the capture zone boundary is the eigenfrequency}

We now need to study the behavior of the vertical eigenvalue as a function of the radius. From our definitions of the quantities involved, it follows that for $a>0$ the product 
$$
\Omega_\perp(r)\epsilon(r)=\Omega(r)-\Omega_\perp(r)+\left(\sigma/m\right)
$$
is a decreasing function of the radius, because such is the difference $\Omega(r)-\Omega_\perp(r)$ [see the basic expressions (\ref{eq:1})]. Thus in this case $\epsilon(r)$ has at most one root (zero) between $r_i$ and $r_{max}$(the outer radius of the disk), since $\Omega_\perp(r)>0$. If  $\epsilon(r_i)<0$, then  $\epsilon(r)$ remains negative in the whole range $r_i<r<r_{max}$, or, equivalently, $\Psi(r)/\omega_*^2(r)>1$ there. However, in such a case we are dealing, by definition, with a g--mode rather than a c--mode.

Hence we have to assume that $\epsilon(r_i)>0$, which converts into a restriction on the eigenfrequency, $\sigma +m\left[\Omega(r_i)-\Omega_\perp(r_i)\right]>0$. Then at any value of the root $r_c$, equation (\ref{eq:9}) gives
\be
\sigma=-m\left[\Omega(r_c)-\Omega_\perp(r_c)\right],\qquad r_i<r_c<r_{-} \; .
\label{eq:17}
\ee
The second inequality here is proved in the following way. The definition of $r_-$ given in section 3 implies $\omega(r_-)=\pm\kappa(r_-)$, with the plus sign for $m>0$ and the minus sign for $m<0$. Hence from the definition (\ref{eq:9}) of the parameter $\epsilon(r)$ we derive, using equation (\ref{eq:2}):
\be
\epsilon(r_-)=-1+\frac{\omega(r_-)}{m\Omega_\perp(r_-)}=-1+\frac{\kappa(r_-)}{|m|\Omega_\perp(r_-)}<0 \; . \label{eq:18}
\ee
So $\epsilon(r)$ is positive at $r=r_i$, negative at $r=r_-$, and the single root $r_c$ of this function is between these two points.
Thus $\Psi(r)/\omega_*^2(r)<1$ for $r_i<r<r_c$, and $\Psi(r)/\omega_*^2(r)>1$ for $r_c<r<r_{-}$. So the c-mode is trapped in $r_i<r<r_c$, and $r_c$ is the outer boundary of the capture zone.

Expression (\ref{eq:17}) for the c-mode eigenfrequency in the case $m^2=1$ has the same structure as the result obtained by the approximate dispersion relation \citep{k90,k93}, with a significant advantage that the value of the radius involved in it is uniquely specified as the right capture zone boundary of the mode in question. Of course, now it is also rigorously derived from the c-mode definition.

Moreover, for $r>r_i(a)\geq1$, $|a|<1$ is smaller than any positive power of $r$. In particular, $a/r^{1/2}\ll 1$, except for $a$ very close to $1$ (where some of the above considerations do not apply, anyway; e.~g., $\epsilon(r_i)\rightarrow\infty$ as $a\rightarrow 1$). Using this fact, from equations (\ref{eq:1}) and (\ref{eq:17}) we obtain
\be
\sigma=-m\frac{2a}{r_c^3}\left[1+{\cal O}\left(\frac{a}{r_c^{1/2}}\right)\right] \; .
\label{eq:19}
\ee
In physical units, the quantity $2a/r_c^3$ is the Lense--Thirring frequency $2GJ/(c^2r_c^3)$. Thus, to lowest order in $a/(r_c)^{1/2}$, {\it the c-mode eigenfrequency $|\sigma|$ is an integer multiple of the Lense--Thirring frequency at $r=r_c$. In particular, the eigenfrequency of the fundamental c--mode ($m^2=j=1$) coincides with the Lense--Thirring frequency at its outer trapping zone boundary.} For $a/r_c^{3/2}<a/r_c^{1/2}\ll 1$, the fundamental c--mode is a low frequency oscillation. 

We can now replace the eigenfrequency $\sigma$ by its dependence on $r_c$ given by equation (\ref{eq:17}), 
\be
\omega(r,r_c)=m\left[\Omega(r)-\Omega(r_c)+\Omega_\perp(r_c)\right] \; ,\quad
\epsilon(r,r_c)=\frac{\left[\Omega(r)-\Omega_\perp(r)\right]-\left[\Omega(r_c)-\Omega_\perp(r_c)\right]}{\Omega_\perp(r)} \; , \label{eq:20}
\ee
and use $r_c$ as a spectral parameter in the radial eigenvalue problem. Its eigenvalue will fix unambiguously the value of both the capture zone boundary $r_c$ and the eigenfrequency $\sigma$. Unlike other recent suggestions regarding the manifestation of the Lense--Thirring effect in accretion disks around black holes \citep{czc}, the value of the radius corresponding to a Lense--Thirring frequency is not a (loosely constrained) parameter, but is predicted by the theory.

Note that for $a<0$ the product $\Omega_\perp(r)\epsilon(r)$ is an increasing function of the radius, so in this case we have to require that $\epsilon(r_i)<0$. However, as seen from equation (\ref{eq:18}), $\epsilon(r_-)$ is still negative, so the single root $r_c$ of $\epsilon(r)$ is greater than $r_-$. In fact, it is also greater than $r_+$, because of the expression similar to equation (\ref{eq:18}), $\epsilon(r_+)=-1-\kappa(r_+)/\left[|m|\Omega_\perp(r_+)\right]<-1$. Hence the capture zone is $r_-<r<r_+$, and again we are dealing with a g--mode rather then a c--mode.  Therefore, {\it there are no c--modes of the studied type for 
counter-rotating disks}. Accordingly, only the case of corotation is investigated in the sequel.

Finally, both the co- and counter-rotating c--modes may have, in principle, a second oscillation domain, $r_-<r<r_+$ and $r_+<r_c<r$, respectively. (Note that the coefficient of $V_r$ in the radial equation (\ref{eq:4}) is also positive there.) These domains, however, are situated at rather large values of the radius (up to infinity in the second case). But expression (\ref{eq:9}) or (\ref{eq:20}) shows that for a given value of $r_c$ and large $r$, $\epsilon(r)\propto r^{3/2}\gg1$, in severe contradiction with the definitive property of the modes in question. Even
$|\epsilon(r_+)|>1$, as seen from the above. In effect, the c--modes of the studied type can exist only in a corotating disk and are trapped in its innermost region.

\subsection{Vertical eigenvalue problem for slightly buoyant disks}

The above theory can be readily extended to the case of a disk with a small buoyancy. It can be characterized by a modified pressure dependence [compare with equation (\ref{eq:3}) and with formula (4.16) from \citet{per}]
\be
 p=p_0(r)(1-y^2)^{\Gamma/(\Gamma-1)}[1-f(r,y)] \; , \label{eq:21}
\ee
where the even in $y$ and slowly varying with $r$ perturbation $f$ is small, $|f(r,y)|\ll 1$. To lowest order this results in the following change of the vertical equation (\ref{eq:8}):
\be
(1-y^2)\frac{d^2V_y}{dy^2} - \left[\frac{2y}{\Gamma -1}+(1-y^2)\delta_1\right]\frac{dV_y}{dy} + \frac{2\omega_*^2}{\Gamma -1}\left[1-(\chi-\delta_2)(1-y^2)\right]V_y = 0 \; ; \label{eq:22}
\ee
where
$$
\delta_1=\delta_1(f)(r,y)=\frac{\Gamma-1}{2\Gamma}\times
$$
$$
\left\{\frac{2\Gamma-1}{\Gamma-1}f^{'}
-\left(\frac{1}{y^{2}}-1\right)\left(f^{'}-yf^{''}\right)-
\frac{1}{\omega^{2}_*}\left[\left(3-\frac{1}{y^{2}}\right)f^{'}+\left(5y+\frac{1}{y^{1}}\right)f^{''}-(1-y^2)f^{'''}\right]\right\};
$$
$$
\delta_2=\delta_2(f)(r,y)=\frac{\Gamma-1}{2y}\left\{f^{'}+
\frac{\Gamma-1}{\omega^2_*\Gamma}\left[\frac{1}{(y^{2}}f^{'}-(\frac{1}{y^{1}}+y)f^{''}+0.5(1-y^2)f^{'''}\right]\right\}
$$

\noindent Here and below the prime means a derivative in $y$, and it is easy to check that the above expressions remain finite at the midplane $y=0$ because $f(r,y)$ is even in $y$. According to the definition (\ref{eq:7}), $\chi(r)=1-\Psi(r)/\omega_*^2(r)$, and it is small for the studied c--mode, so there are three possible situations depending on how the new small corrections compare to it. If the buoyancy-induced terms $\delta_1$ and $\delta_2$ are much smaller than $\chi$, i.~e., than our basic small parameter $\epsilon$ from the equality (\ref{eq:9}), then nothing changes in the main approximation. In the opposite case the very existence of the modes we are looking for is questionable, and the whole theory should be changed essentially. Thus the only interesting possibility is when both perturbations are of the same order. This assumption can always be made for a given degree of buoyancy. In such a case by the same standard technique of discrete spectrum perturbations we obtain, instead of the relation (\ref{eq:10}), the form
\be
\chi=\chi(r,a,m,j,\Gamma)=\chi_1\epsilon(r)+\tilde{\chi_1}+\dots,
\label{eq:23}
\ee
where $\chi_1(m,j,\Gamma)$ is still given by expression (\ref{eq:14}) and $m$, $j$ and $\Gamma$ are related by the selection rule (\ref{eq:12}). We obtain
\be
\tilde{\chi_1}=\tilde{\chi_1}(r,a,m,j,\Gamma)=\frac
{\big\langle (1-y^2)\delta_2(f)V^0,V^0\big\rangle-(\Gamma-1)(2m^2)^{-1}\big\langle\delta_1(f)(1-y^2) [V^0]^{'},V^0\big\rangle}{\big\langle(1-y^2) V^0,V^0\big\rangle} \; . \label{eq:24}
\ee
When calculating this, one should replace $\omega^{2}_*$ by $m^2$ in the above expression for $\delta_2$. The unperturbed eigenfunction $V^0=V^0_j(y)$ and the scalar product $\langle\cdot,\cdot\rangle$ are defined in section~4.1.

As for the radial behavior of the vertical eigenvalue studied in section 4.2 for zero buoyancy, it might not change qualitatively due to the new term $\tilde{\chi_1}$, in which case only the value of $r_c$, specified now by the equation $\chi(r_c)=0$ instead of $\epsilon(r_c)=0$, changes somewhat. Otherwise the whole mode capture zone changes. For instance, it turns into the interval $(r_i,\,r_-)$ if $\chi(r)$ remains positive throughout it.

To understand the results of such possible changes, as an example let us treat a buoyancy perturbation which might be considered a typical one for thin disks,
\be
f(r,y)=\xi(r)y^2 \; . \label{eq:25}
\ee
The buoyancy corrections then become (for $m^2=1$)
\be
\delta_1=\xi(r)\frac{7-6\Gamma}{\Gamma}y \; , \qquad 
\delta_2=\xi(r)\frac{\Gamma-1}{\Gamma} \; , \label{eq:26}
\ee
and the resulting expressions simplify greatly. In particular, for the fundamental c-mode the formula (\ref{eq:15}) generalizes to
\be
\Psi(r)/\omega_*^2(r)=1-\left[(3\Gamma-1)\epsilon(r)+\frac{(6\Gamma-5)(\Gamma-1)}{2\Gamma^2}\xi(r)\right]+\dots,\qquad (m^2=j=1) \; .
\label{eq:27}
\ee
It is clear that the radial behavior depends on the value of $\Gamma$ and on how $\xi(r)$ compares to $\epsilon(r)$.
We consider only non-buoyant disks in the sequel.

\section{Radial Eigenvalue Problem: Determining the Capture Zone and the Eigenfrequency}

\subsection{Formulation of the problem}

Recall that we are now dealing with the case $a>0$, in which case the c--modes are trapped near the inner edge of the disk, $r_i<r<r_c$. We require the modes to decay within $r_c<r<r_-$; this requirement provides the boundary condition at $r=r_c$ for the radial equation (\ref{eq:5}). The boundary condition at the inner edge is not easy to specify, so we employ a general linear combination,
\be
\cos{\theta}\,\frac{dV_r}{dr}-\sin{\theta}\, V_r\biggl|_{r=r_i}=0 \; , \qquad
0\leq\theta\leq\frac{\pi}{2} \; ; \label{eq:28}
\ee
keeping the lack of knowledge parameterized by ${\theta}\in\left[0,\,\pi/2\right]$. 

We introduce a new independent variable
\be
\tau(r)=\int_{r_i}^r\left[\omega^2(r^{'})-\kappa^2(r^{'})\right]\,dr^{'} \; , \qquad \tau_c\equiv\tau(r_c) \; , \label{eq:29}
\ee
which increases monotonically from zero to $\tau_c$ when $r$ goes from $r_i$ to $r_c$, so the inverse transformation is well defined. The new variable allows us to rewrite equation (\ref{eq:5}) in the WKB form:
\be
\frac{d^2 V_r}{d\tau^2} +\frac{\chi_1\epsilon\alpha^2}{\omega^2-\kappa^2}V_r = 0 \; .
\label{eq:30}
\ee
The boundary condition (\ref{eq:28}) in terms of $\tau$ becomes:
\be
\omega^{2}(r_i)\cos{\theta}\,\frac{dV_r}{d\tau}-\sin{\theta}\, V_r\biggl|_{\tau=0}=0 
\; . \label{eq:31}
\ee
The coefficient in front of $V_r$ in equation (\ref{eq:30}) is obtained from its general expression (\ref{eq:15}) to lowest order in $\epsilon$. For convenience, we assign a special notation to it,
\be
Q(\tau)=\frac{\chi_1\epsilon\alpha^2}{\omega^2-\kappa^2} \; . \label{eq:32}
\ee
All factors here can be considered as functions of $\tau$ via the dependence $r=r(\tau)$ inverse to the function (\ref{eq:29}).

We now need to describe the behavior of $\alpha(r)$, which is determined by the fact that this function is essentially the inverse speed of sound at the midplane, $\alpha(r)\propto c_s^{-1}(r,0)$. Since the pressure at the inner edge vanishes in a typical disk model [$p\propto (r-r_i)^{k}$, with $0 < k \leq 1$] if the torque does, we obtain $\alpha(r)\propto (r-r_i)^{-\mu}$ with $\mu=(\Gamma-1)k/2\Gamma$. If even a small torque is applied to the disk at the inner edge, which is not unlikely in practice, the pressure and the speed of sound become nonzero at $r=r_i$. Hence $\alpha(r)$ loses its weak singularity there. This situation is included by assigning $\mu=0$.

It is also known that the speed of sound drops as some power at large radii. Incorporating all this information, we are now able to specify the structure of the function $\alpha(r)$ as
\be
\alpha(r)=\gamma(r)\frac{r^{\mu+\nu}}{(r-r_i)^{\mu}} \; , \qquad 
0 \leq \mu < 1/2 \; , \quad \nu\geq 0 \; , \label{eq:33}
\ee
where $\gamma(r)$ is some function bounded from above and away from zero, varying slowly with radius, and tending at infinity to a limit $\gamma(\infty) \equiv \gamma_\infty$. A typical value of $\mu$ corresponding to a simple root ($k=1$) of the pressure at $r_i$ and the value $\Gamma=5/3$ is $\mu=2/5$. It is also true that  $\nu<1$ in most disk models. 

Our radial eigenvalue problem is thus completely defined. Therefore we proceed to determine its eigenvalues.

\subsection{WKB solution of the radial eigenvalue problem}

We apply the WKB procedure to the eigenvalue problem specified by equations (\ref{eq:30}), (\ref{eq:31}), and the decay condition for $\tau>\tau_c$. This seems to be the only effective analytical approach. It is validated partly by the fact that despite $\epsilon(r)$ being small, the coefficient $Q(\tau)$ is still rather large everywhere except close to $\tau_c$ because $\alpha(r)$ is large. This follows from equation (\ref{eq:33}) since $\gamma\sim c/c_s$, with $c$ the speed of light and $c_s(\ll c)$ the speed of sound [for estimates of the WKB approximation accuracy see \citet{olv}, for example]. Thus we represent the solution as
\be
V_r\,\propto\, Q^{-1/4}(\tau)\,\cos\left[\Phi(\tau)-\Phi_c\right] \; ,\qquad \Phi(\tau)=\int_0^\tau Q^{1/2}(\tau^{'})\,d\tau^{'} \label{eq:34}
\ee 
in the whole capture zone $0<\tau<\tau_c$, except small vicinities of its boundaries $\tau=0$ and $\tau=\tau_c$. Here $\Phi_c$ is a constant phase, so far unknown. 

The right boundary $\tau=\tau_c$ proves to be a turning point of equation (\ref{eq:30}). As shown in section 4.2, $\epsilon(r)$, and hence $Q(\tau)$, change sign there, being positive on the left of $\tau_c$ and negative on the right of it, with $\epsilon(r_c)=Q(\tau_c)=0$. Near $\tau_c$ we have $Q(\tau)=-q_c\left(\tau-\tau_c\right)+...$, with $q_c=-Q^{'}(\tau_c)>0$. Hence equation (\ref{eq:30}) reduces to the Airy equation, whose solution decaying for $\tau>\tau_c$ is given by the Airy function of the first kind,
\be
V_r\,\propto Ai\left(q^{1/3}_c\left(\tau-\tau_c\right)\right)\; . \label{eq:35}
\ee 
Matching in the usual way the negative infinity asymptotics of the function (\ref{eq:35}) with the $\tau\rightarrow\tau_c-0$ limit of equation (\ref{eq:34}), we obtain the following expression for the phase $\Phi_c$ (where $n$ is an integer related to the number of radial nodes of the mode):
\be
\Phi_c=\int_0^{\tau_c} Q^{1/2}(\tau)\,d\tau+\frac{\pi}{4}-\pi n \; . \label{eq:36}
\ee

Now, close to the left boundary $\tau=0$, according to equations (\ref{eq:32}), (\ref{eq:33}), and (\ref{eq:29}), the coefficient $Q(\tau)$ has a weak singularity, 
\be
Q(\tau)=q_i\tau^{-2\mu}+...,\quad\tau\rightarrow+0 \; ; \qquad 
q_i(r_c,a)\equiv\chi_1\gamma^2(r_i)r_i^{2(\mu+\nu)}\left|\omega(r_i)\right|^{2(1-\mu)}\epsilon(r_i) > 0 \; , \label{eq:37}
\ee
and equation (\ref{eq:30}) becomes
$$
\frac{d^2 V_r}{d\tau^2} +\frac{q_i}{\tau^{2\mu}}V_r = 0 \; .
$$
Its exact solution satisfying the boundary condition (\ref{eq:31}) is given in terms of Bessel functions as
\be
V_r\,\propto \tau^{1/2}\left[\sin\theta\frac{\left(\lambda/2\right)^{-\zeta}}{\Gamma_-}J_{\zeta}\left(\lambda\tau^{1-\mu}\right)+
\omega^2(r_i)\cos\theta\frac{\left(\lambda/2\right)^{\zeta}}{\Gamma_+}J_{-\zeta}\left(\lambda\tau^{1-\mu}\right) \right] \; , \label{eq:38}
\ee 
where for brevity we have introduced the notations
\be
\lambda(r_c,a)\equiv\frac{q_i^{1/2}(r_c,a)}{1-\mu} \; , \quad
\zeta\equiv\frac{1}{2(1-\mu)} \; , \quad
\Gamma_{\pm}\equiv\Gamma(1\pm\zeta) \; . \label{eq:39}
\ee
[$\Gamma(z)$ is the Euler gamma-function.] Using the large argument expressions of the Bessel functions, it is straightforward to write the `far field' ($\lambda\tau^{1-\mu}\gg1$) asymptotics of the solution (\ref{eq:38}) as
\be
V_r\,\propto \tau^{\mu/2}\cos\left(\lambda\tau^{1-\mu}-\frac{\pi}{4}+\Phi_i\right) 
\; , \label{eq:40}
\ee 
where
\be
\Phi_i(r_c,a)\equiv\arctan\left(\frac{Z_-}{Z_+}\tan\frac{\pi\zeta}{2}\right) , \quad 
Z_\pm(r_c,a) \equiv\frac{\Gamma_-}{ \Gamma_+}\omega^2(r_i) \left(\frac{\lambda}{2}\right)^{2\zeta}\cos\theta\pm\sin\theta \; . \label{eq:41}
\ee 

The last step is to match the asymptotic expression (\ref{eq:40}) with the $\tau\rightarrow+0$ limit of the WKB solution (\ref{eq:34}). In view of the expression (\ref{eq:37}), the latter is easily found to be
$$
V_r\,\propto \tau^{\mu/2}\cos\left(\lambda\tau^{1-\mu}-\Phi_c\right) \; ,
$$ 
so the match requires that
$$
\Phi_c+\Phi_i=\frac{\pi}{4} \; .
$$
Employing the expression (\ref{eq:36}), we then arrive at the desired eigenvalue equation
\be
\Phi_i+\int_0^{\tau_c} Q^{1/2}(\tau)\,d\tau=\pi n \; . \label{eq:42}
\ee
For a given set of all the parameters, including $a$ and $n$, any solution $\tau_c$ of this equation specifies the corresponding c-mode boundary $r_c$ according to the relations (\ref{eq:20}) and (\ref{eq:29}). The eigenfrequency $\sigma$ of the mode is then found from $r_c$ by the general formula (\ref{eq:17}). So we now study the roots $\tau_c$ of equation (\ref{eq:42}).

\subsection{The c--mode eigenfrequency spectrum is at most finite}

It is now convenient to work directly with $r_c$ rather than with $\tau_c$. So first of all we rewrite the eigenfrequency equation (\ref{eq:42}) in the following form:
\be
{\cal F}(r_c,a)\equiv\Phi(r_c,a)+I(r_c,a)= n \; , \qquad n=0,1,2,\dots \; ; \label{eq:43}
\ee
where
\be
\Phi(r_c,a)\equiv\frac{1}{\pi}\Phi_i(r_c,a) \; , \label{eq:44}
\ee
\be
I(r_c,a)\equiv\frac{1}{\pi}\int_{r_i}^{r_c}\frac{\gamma(r)r^{\mu+\nu}}{(r -r_i)^{\mu}}\sqrt{\chi_1\epsilon(r,r_c)\left[m^2\Omega^2_\perp(r)-\kappa^2(r)\right]}\,dr \; , \qquad \label{eq:45}
\ee
assuming that $0\leq\mu<1/2$ and $\quad 0\leq\nu < 1$.
We returned to the integration over $r$ in equation (\ref{eq:42}), and used the definition (\ref{eq:32}) of $Q(\tau)$ and the expression (\ref{eq:33}) for $\alpha(r)$. We have also replaced $\omega^2(r)$ by $m^2\Omega^2_\perp(r)$, which is correct to lowest order in $\epsilon(r)$ according to the definition (\ref{eq:9}) of the latter parameter. [Within the studied approximation, the same should be done with $\omega^2(r_i)$ involved in $\Phi_i(r_c,a)$; see the second of the equalities (\ref{eq:41}).] 

The integer $n$ in the equation (\ref{eq:43}) should not be negative for the following reason. It is straightforward to see from equation (\ref{eq:41}) that in any case $|Z_-/Z_+|\leq1$ (recall that $0\leq\theta\leq\pi/2$). Hence
\be
|\Phi(r_c,a)|\leq\frac{1}{\pi}\arctan\left(\tan\frac{\pi\zeta}{2}\right)\leq\frac{\zeta}{2}=\frac{1}{4(1-\mu)}<\frac{1}{2} \; . \label{eq:46}
\ee
Obviously, $I(r_c,a)$ is non-negative for all $r_c\geq r_i$, therefore the whole left hand side of equation (\ref{eq:43}) is always larger than $-0.5$ and thus cannot equal the right hand side with $n=-1,-2,\dots$.

For a given set of parameters [including $m^2$ and $j$ satisfying the selection rule (\ref{eq:12})] and $0\leq a\leq1$ let us now study the roots of equation (\ref{eq:43}). First of all, $\epsilon(r)\equiv0$ for $a=0$; thus $I(r_c,0)=0$ and no roots can be found. There is no c-mode for a strictly non-rotating disk. So the  rotation parameter interval is $0<a<1$. Then it is easy to see from the expression (\ref{eq:20}) for $\epsilon$ and the inequalities (\ref{eq:2}) that for any $r\geq r_i$,
\be
\epsilon(r,r_c)<\epsilon_{\infty}(r)\equiv\lim_{r_c\rightarrow\infty}\epsilon(r,r_c)=
\frac{\Omega(r)-\Omega_\perp(r)}{\Omega_\perp(r)} \; . \label{eq:47}
\ee
Therefore there exists a finite limit
\bea
I(\infty,a) & \equiv & \lim_{r_c\rightarrow\infty}I(r_c,a) = \frac{1}{\pi}\int_{r_i}^{\infty}\frac{\gamma(r)r^{\mu+\nu}}{(r-r_i)^{\mu}} \sqrt{\chi_1\epsilon_\infty(r)\left[m^2\Omega^2_\perp(r)-\kappa^2(r)\right]}\,dr \nonumber \\
 & = & \frac{\sqrt\chi_1}{\pi}\int_{r_i}^{\infty}
\frac{\gamma(r)r^{\mu+\nu}}{(r-r_i)^{\mu}}\sqrt{\frac{
\left[\Omega(r)-\Omega_\perp(r)\right]\left[m^2\Omega^2_\perp(r)-\kappa^2(r)\right]}{\Omega_\perp(r)}}\,dr < \infty \; . \label{eq:48}
\eea
Indeed, the integral (\ref{eq:48}) converges at the upper limit because its integrand is $\propto r^{\nu-11/4}$ ($m^2=1$) and $\propto r^{\nu-9/4}$ ($m^2>1$) when $r\rightarrow\infty$ [see the expressions (\ref{eq:1})], and $\nu$ is always less than one. For the special case $a=1$, $\Omega_\perp(r)$ has a simple root at $r=r_i=1$, so the singularity at the lower limit becomes $(r-r_i)^{-(\mu+1/2)}$. However, the integral still converges since $\mu<1/2$. 

Hence $I(r_c,a)$ is a bounded function of $r_c$ on the semiaxis, and there exists
\be
I_m(a)\equiv \sup_{r_c\geq r_i}I(r_c,a) \; , \qquad 0<I_m(a)<+\infty\; .\label{eq:49}
\ee
By the inequality (\ref{eq:46}), we have already shown the existence of
\be
\Phi_m(a)\equiv \sup_{r_c\geq r_i}\Phi(r_c,a) \; , \qquad 
\Phi_m(a)\leq 0.25(1-\mu)^{-1} \; . \label{eq:50}
\ee
This implies that the left hand side of equation (\ref{eq:43}) is a bounded function of $r_c$:
\be
{\cal F}_m(a)\equiv \sup_{r_c\geq r_i}{\cal F}(r_c,a) \; , \qquad 
-0.25(1-\mu)^{-1}<{\cal F}_m(a)<+\infty \; . \label{eq:51}
\ee
But then {\it equation} (\ref{eq:43}) {\it has no roots for all integers $n>{\cal F}_m(a)$}, so {\it only c-modes with the radial numbers $0\leq n\leq{\cal F}_m(a)$ can exist}. In other words, {\it for any value of $a$, $0< a\leq1$, only c-modes with the radial numbers $n=0,1,2,\dots,N(a)$ can exist, with the integer $N(a)$ satisfying} ${\cal F}_m(a)-1<N(a)\leq{\cal F}_m(a)$.
Along with the analytic dependence of ${\cal F}(r_c,a)$ on $r_c$, this also shows that {\it the set of roots of equation} (\ref{eq:43}) {\it for each admissible value of $n$ is discrete}; but we actually can do much better than that. 

Namely, we first recall from section 4.2 that for any positive value of $a$, the difference $\left[\Omega(r)-\Omega_\perp(r)\right]$ is a {\it decreasing positive function of $r$}. Then the equalities (\ref{eq:20}), containing the term $-\left[\Omega(r_c)-\Omega_\perp(r_c)\right]$, show us that both $|\omega(r,r_c)|$ and $\epsilon(r,r_c)$ {\it are increasing functions of $r_c$ for any $r\geq r_i$}. 
This allows us to conclude that {\it for any $0< a\leq1$ the left hand side ${\cal F}(r_c,a)$ of equation} (\ref{eq:43}) {\it is an increasing function of $r_c$}. 

Indeed, from the equalities (\ref{eq:41}) and definitions (\ref{eq:39}) and (\ref{eq:37}) we see that for $\theta\not=0,\pi/2$, the phase $\Phi(r_c,a)$ depends on $r_c$ only through the combination $|\omega(r_i,r_c)|^3\left[\epsilon(r_i,r_c)\right]^\zeta$, and the dependence is monotonically increasing. Since this combination is an increasing function of $r_c$, so is the function $\Phi(r_c,a)$ itself. In particular, its maximum is 
\be
\Phi_m(a)=\Phi(\infty,a)\equiv\lim_{r_c\rightarrow\infty}\Phi(a,r_c) \label{eq:52}
\ee
[see equation (\ref{eq:50})], whose value can easily be calculated explicitly  according to equations (\ref{eq:41}), (\ref{eq:39}), (\ref{eq:37}) and (\ref{eq:20}). For the cases $\theta=0,\pi/2$, i.~e., the boundary condition (\ref{eq:28}) of the second and first kind, respectively, $\Phi$ is independent of $r_c$ and $a$,
\be
\Phi(r_c,a)=\pm\frac{1}{4(1-\mu)}={\rm constant} \; , \label{eq:53}
\ee
and the relation (\ref{eq:52}) is still formally true.

Next, the definition (\ref{eq:45}) shows that
$$
\frac{\partial I(r_c,a)}{\partial r_c}=\frac{1}{2\pi}\int_{r_i}^{r_c}
\frac{\gamma(r)r^{\mu+\nu}}{(r-r_i)^{\mu}}\,
\sqrt{\frac{\chi_1\left[m^2\Omega^2_\perp(r)-\kappa^2(r)\right]}{\epsilon(r,r_c)}}\,
\frac{\partial \epsilon(r,r_c)}{\partial r_c}\, dr>0 \; ,
$$
so $I(r_c,a)$ is also an increasing function of $r_c$, with the maximum value [see (\ref{eq:49})]
\be
I_m(a)=I(\infty,a) \label{eq:54}
\ee
given by expression (\ref{eq:48}). Thus ${\cal F}(r_c,a)=\Phi(r_c,a)+I(r_c,a)$  increases with $r_c$ to its maximum value at 
$r_c=\infty$,
\be
{\cal F}_m(a)=\Phi_m(a)+I_m(a)=\Phi(\infty,a)+I(\infty,a)={\cal F}(\infty,a)\; .
\label{eq:55}
\ee
Our last observation regarding ${\cal F}(r_c,a)$ is that
\be
-0.25(1-\mu)^{-1}\leq{\cal F}(r_i,a)=\Phi(r_i,a)\leq0.25(1-\mu)^{-1}<n \; , \qquad n=1,2,3,\dots\; . \label{eq:56}
\ee

The implication of the established properties of ${\cal F}(r_c,a)$ is clear and most important. Namely, {\it for any given $a$ ($0<a<1$), and any $n=1,2,3\dots,N(a)$, with the integer $N(a)$ satisfying
\be
{\cal F}_m(a)-1<N(a)<{\cal F}_m(a)\; , \label{eq:57}
\ee
there exists exactly one root $r_c^{(n)}(a)\in (r_i,\,\infty)$ of the increasing function $\left[{\cal F}(r_c,a)-n\right]$, i.~e., of equation} (\ref{eq:43}). {\it Moreover, if 
\be
{\cal F}(r_i,a)=\Phi(r_i,a) < 0 \; ,\qquad
{\cal F}_m(a)=I(\infty,a)+\Phi(\infty,a)>0 \; , \label{eq:58}
\ee
then in addition there exists a root $r_c^{(0)}(a)\in (r_i,\,\infty)$ of equation} (\ref{eq:43}) {\it with $n=0$. All the roots prove to be numbered in an increasing order,}
\be
r_i < \left[ r_c^{(0)}(a) \right] < r_c^{(1)}(a) < r_c^{(2)}(a) < \dots<r_c^{(N)}(a) \; , \qquad N = N(a) \; .  \label{eq:59}
\ee
Turning now to the eigenfrequencies as defined through the corresponding capture zone boundaries by the general formula (\ref{eq:17}), we can reformulate our result as follows. 

{\it For any given $a$, $0< a\leq1$ the spectrum of the studied c-modes consists of a finite number of eigenfrequencies $\sigma^{(n)}(a),\,n=[0],1,2,3\dots,N(a)$; with $N(a)$ specified by the inequalities} (\ref{eq:57}), {\it and $\sigma^{(0)}(a)$ existing only under the conditions} (\ref{eq:58}). {\it The eigenfrequencies are negative for $m>0$ and positive for $m<0$, and their magnitudes are always ordered as}
\be
|m|\left[\Omega(r_i)-\Omega_\perp(r_i)\right] > \left[|\sigma^{(0)}(a)|\right] > |\sigma^{(1)}(a)|>|\sigma^{(2)}(a)|>\dots>|\sigma^{(N)}(a)|>0\; ,\quad\label{eq:60}
\ee
with $N=N(a)$. Note that this is true for a given value of the parameter $\Gamma$ and any angular and vertical numbers $m$ and $j$ satisfying the selection rule (\ref{eq:12}), so that the complete notation for the eigenfrequency is $\sigma^{(n)}_{mj}(a)$. 

Note also that if ${\cal F}_m(a)<1$, then we have $N(a)=0$ by the right inequality in equation (\ref{eq:57}), and the only c-mode that can possibly exist is that with $n=0$. On the other hand, the number $N(a)$ is bounded for all values of $a\in[0,\,1]$ by the maximum value 
\be
{\cal F}_{max}\equiv\max_{0\leq a\leq1}{\cal F}_m(a)={\cal F}_m(a^*)\; , \qquad a^*\in[0,\,1]\; , \label{eq:61}
\ee
$N(a)<{\cal F}_{max}$. So the total number of different c-modes with given angular and vertical numbers is less than ${\cal F}_{max}+1$ for the whole family of corotating disks which differ only by the value of the angular momentum of the central body. If in particular ${\cal F}_{max}<1$, which can happen if the speed of sound at the midplane is everywhere sufficiently large inside the disk, then at most one c-mode with $n=0$ may be excited in any disk of the family.

\subsection{Dependence of the spectrum on the rotation parameter: the cutoff values of $a$ ($n\geq1$)}

We now change our point of view: instead of studying all the roots for a given $a$, we specify some $n\geq 1$ and investigate how the trapping radius $r_c^{(n)}(a)$, determined by
\be
{\cal F}\left(r_c^{(n)},a\right)=\Phi(r_c^{(n)},a)+I(r_c^{(n)},a)=n \; ,\label{eq:62}
\ee
behaves as a function of $a$. An intuitive expectation is that it monotonically decreases with $a$, but that might not be generally true. The differentiation of equation (\ref{eq:62}) gives
$$
\frac{dr_c^{(n)}}{da}=
-\frac{\left(\partial{\cal F}/\partial a\right)}
{\left(\partial{\cal F}/\partial r_c\right)}\Biggl|_{r_c=r_c^{(n)}(a)} \; ,
$$
and, although the denominator here is positive, it is difficult to establish that the sign of the numerator is also always positive. Even the sign of $\partial{I}/\partial a$ is not always obvious, because of many dependences on $a$ in the expression (\ref{eq:45}): the quantities $\epsilon$, $\Omega_\perp$, $\kappa$, and  $r_i$ all depend on $a$. 

Nevertheless, $I(\infty,a)$ goes to zero when $a\rightarrow+0$ (since $\Omega_\perp\rightarrow\Omega$), so $I(r_c,a)<I(\infty,a)$ goes to zero uniformly in $r_c\geq r_i$ as well. In particular, $I(r_c,a)<n-1/2$ for small enough positive values of $a$. Since the first term on the left of equation (\ref{eq:43}) is less than $1/2$ by equation (\ref{eq:46}), {\it equation} (\ref{eq:43}) {\it has no root at all for such values of $a$}. Therefore $r_c^{(n)}(a)$ does not exist in some positive vicinity of $a=0$.

Suppose now that $r_c^{(n)}(a)$ does exist for some $a_*>0$; that is, $n\leq N(a_*)$. By making the value of $a$ smaller and smaller  we would eventually find $r_c^{(n)}(a)$ growing, to make up for the reduced value of $I(r_c^{(n)}(a),a)<I(\infty,a)\rightarrow+0$. Recall that for a given $a$, $I(r_c,a)$ is a growing function of $r_c$. At a certain value $a_n>0$, only $r_c^{(n)}(a_n)=\infty$ would satisfy equation (\ref{eq:62}),
\be
{\cal F}\left(\infty,a_n\right)=\Phi(\infty,a_n)+I(\infty,a_n)=n \; . \label{eq:63}
\ee
For $a<a_n$, no value of the capture zone radius would satisfy it, so $r_c^{(n)}(a)$ no longer exists.

Therefore we have proved that {\it for any admissible radial mode number $n\geq1$ there exists a cutoff value $a_n>0$ such that $r_c^{(n)}(a)$, and thus the corresponding c-mode, exists for $a>a_n$, and does not exist for $0\leq a\leq a_n$. When $a\rightarrow a_n+0$, the capture zone boundary $r_c^{(n)}(a)\rightarrow\infty$ and the eigenfrequency $\sigma^{(n)}(a)\rightarrow0$. The cutoff value is determined by equation} (\ref{eq:63}). {\it Evidently, $a_n$ increases with $n$,
$$
0<a_1<a_2<\dots \; .
$$
Also, the maximum radial mode number $N(a)$ defined by the inequalities} (\ref{eq:57}) {\it is a piecewise constant increasing function with unit jumps at $a=a_n$}.

An approximate expression for the cutoff values can be obtained from equation (\ref{eq:63}) if they are small enough, so that $a_n/r_i^{3/2}(a_n)\ll 1$. We find that
\be
a_n\approx\left[\frac{n+0.25/(1-\mu)}{I}\right]^2,\; 
a_n^{1/2}\ll\sin\theta > 0 \; ; \quad 
a_n\approx\left[\frac{n-0.25/(1-\mu)}{I}\right]^2,\; \theta = 0 \; ;
\label{eq:64}
\ee
where
\be
I=\frac{\sqrt{2\chi_1}}{\pi}\int_{6}^{\infty}
\frac{\gamma(r)r^{\mu+\nu-9/4}}{(r-6)^{\mu}}\sqrt{m^2-1+\frac{6}{r}}\,dr \; , \qquad
I(\infty,a)=\sqrt{a}I+\dots \; . \label{eq:65}
\ee
We have used the fact that $r_i(a)\approx r_i(0)=6$ for small $a$, along with the useful asymptotic expressions implied by equations (\ref{eq:20}) and (\ref{eq:1}):
\bea
\epsilon(r,r_c) & = & 2ar^{3/2}\left(\frac{1}{r^3}-\frac{1}{r_c^3}\right)+\dots\; , \qquad \epsilon(r,\infty) = \frac{2a}{r^{3/2}}+\dots \; ; \label{eq:66} \\
\omega(r,r_c) & = & m\left[\frac{1}{r^{3/2}}-a\left(\frac{1}{r^3}+\frac{2}{r_c^3}\right)+\dots\right] \; , \qquad
m^2\Omega_\perp^2-\kappa^2 = m^2-1+\frac{6}{r}+\dots \; . \label{eq:67}
\eea
The latter, in turn, allow us to obtain, from equations (\ref{eq:44}), (\ref{eq:41}), (\ref{eq:39}), and (\ref{eq:37}), the results
\be
\Phi(r_c,a)=-0.25(1-\mu)^{-1}+{\cal O}\left(a^{1/2(1-\mu)}\right),\; \theta\not= 0 \; ; \quad
\Phi(r_c,a)=0.25(1-\mu)^{-1}, \; \theta = 0 \; . \label{eq:68}
\ee
These relations, for $r_c=\infty$, are also used to derive formulas (\ref{eq:64}).
For the expressions (\ref{eq:64}) to be applicable, one essentially requires $I\gg1$. 

The last item on our agenda for c-modes with $n\geq 1$ is to establish the behavior of $r_c^{(n)}(a)$ and $\sigma^{(n)}(a)$ near the cutoff value $a_n$. Asymptotic calculations based on equations (\ref{eq:62}), (\ref{eq:63}), (\ref{eq:20}), and (\ref{eq:19}) allow us to find the results, which differ slightly depending on the angular number of the mode. 

\medskip
\noindent a) {\it Fundamental c-mode ($m^2=j=1$)}. In this case we obtain, for $a\rightarrow a_n+0$:
\be
r_c^{(n)}(a)\approx C_n\left(a-a_n\right)^{-\frac{1}{7/4-\nu}} , \quad
C_n = \left\{\frac{\sqrt{12a_n\left(3\Gamma-1\right)}\gamma_\infty}
{\left(7/4-\nu\right)\left[\Phi^{'}(\infty,a_n)+I^{'}(\infty,a_n)\right]}
\right\}^{\frac{1}{7/4-\nu}} \; ; \label{eq:69}
\ee
where prime denotes the derivative in $a$, $\gamma_\infty$ is defined in section 5.1 [see equation (\ref{eq:33})], and the expression (\ref{eq:16}) has been used. According to equation (\ref{eq:19}), for the eigenfrequency near its cutoff we have
\be
\sigma^{(n)}(a)=\mp\frac{2a}{\left[r_c^{(n)}(a)\right]^3}\approx 
\mp \frac{2a_n}{C_n^{3}}\left(a-a_n\right)^{\frac{3}{7/4-\nu}} \; ; \label{eq:70}
\ee
and the minus (plus) sign is taken for $m=+1(-1)$.

\medskip
\noindent b) {\it C-modes with higher axial mode numbers ($m^2>1$)}. Now we find (near the cutoff $a\rightarrow a_n+0$):
\be
r_c^{(n)}(a)\approx D_n\left(a-a_n\right)^{-\frac{1}{5/4-\nu}} , \quad
D_n = \left\{\frac{\sqrt{2a_n\left(m^2-1\right)\chi_1}\gamma_\infty}
{\left(5/4-\nu\right)\left[\Phi^{'}(\infty,a_n)+I^{'}(\infty,a_n)\right]}
\right\}^{\frac{1}{5/4-\nu}} \; ; \label{eq:71}
\ee
with the value of $\chi_1$ given by equation (\ref{eq:14}). The corresponding eigenfrequency becomes
\be
\sigma^{(n)}(a)=-\frac{2ma}{\left[r_c^{(n)}(a)\right]^3}\approx 
- \frac{2ma_n}{D_n^{3}}\left(a-a_n\right)^{\frac{3}{5/4-\nu}} \; . \label{eq:72}
\ee
Recall from formula (\ref{eq:53}) that $\Phi^{'}(\infty,a)\equiv0$ for $\theta=\pi/2,\,0$; that is, for boundary conditions of the first and second kind at the inner edge of the disk. Otherwise, the expressions for the coefficients $C_n$ and $D_n$ can also be made more explicit if $a_n$ is small.

Finally, we would like to say a few words about the behavior of the studied c-modes in the limit of very rapidly corotating disks, $a\rightarrow1-0$. It turns out that the basic assumption of our approach becomes invalid in this limit, because the (assumed small) parameter $\epsilon$ acquires large values at the inner edge of the disk, reaching eventually to infinity when $a=1$ [unless the capture zone shrinks, i.~e., unless $r_c^{(n)}(a)\rightarrow r_i(a)\rightarrow1+0$ for $a\rightarrow1-0$]. That happens due to the fact that the denominator of the expression (\ref{eq:9}) or (\ref{eq:20}) for $\epsilon$ goes to zero at the inner edge, $\Omega_\perp(r_i=1)=0$ for $a=1$. However, these misfortunes occur very close to $a=1$, namely, for $a > a_\perp\cong 0.953$, at which value $\Omega_\perp(r)$ as a function of $r$ becomes non-monotonic by developing a maximum at some radius close to the innermost one. Therefore our formulas and conclusions remain valid for $0<a\leq a_\perp$, almost all the way up to the black hole limit $a=1$.

\subsection{C--mode for $n=0$}

Turning to the mode with the radial mode number $n=0$, we rewrite the eigenvalue equation (\ref{eq:43}) as
\be
-\Phi(r_c,a) = I(r_c,a) \; , \qquad n=0 \; . \label{eq:73}
\ee
Since $I(r_c,a)$ is positive, $\Phi(r_c,a)$ must be negative for the mode to exist. This condition, with the use of equations (\ref{eq:44}), (\ref{eq:41}), (\ref{eq:39}), and (\ref{eq:37}), translates into
\be
\left[
0.25\chi_1(1-\mu)^{-1}\gamma^2(r_i)r_i^{2(\mu+\nu)}\epsilon(r_i,r_c)
\right]^{\frac{1}{1-\mu}}\left|\omega(r_i,r_c)\right|^3 < \frac{\Gamma_+}{\Gamma_-}\tan\theta \; . \label{eq:74}
\ee
Thus, the $n=0$ c-mode does not exist at all if $\theta=0$, i.~e., for the boundary condition of the second kind at the inner edge of the disk. We consider now only positive values of $\theta$. For such values, however, the mode is absent if the parameter $a$ is small enough, because the right hand side of equation (\ref{eq:73}) is ${\cal O}(a^{1/2})$ by equation (\ref{eq:65}), while the left hand side approaches $0.25(1-\mu)^{-1}$ by equation (\ref{eq:68}). A sufficient condition for the mode to exist at least for some values of $a$ is evidently 
$$
\max_{0<a\leq0.95}I(\infty,a)>\frac{1}{4(1-\mu)} \; .
$$

If it exists, the $n=0$ mode thus has a cutoff value $a_0$, as the modes with other radial numbers do, and this cutoff is the smallest one. Its approximate magnitude is given by the first of equations (\ref{eq:64}) with $n=0$,
$$
a_0\approx\frac{1}{16(1-\mu)^2I^2} \; ,
$$
and the behavior of the capture zone boundary and the eigenfrequency near the cutoff is described by either expressions (\ref{eq:69}), (\ref{eq:70}) or (\ref{eq:71}), (\ref{eq:72}), with $n$ set to zero. 

Therefore there are no c-modes of the studied type when the disk rotation is sufficiently slow. Namely, {\it if the boundary conditions at the inner edge are of the second kind, then there are no c-modes for $0\leq a < a_1$. For all other boundary conditions, the same is true in a smaller interval of rotation parameter values, $0\leq a < a_0 < a_1$.}

\bigskip

This concludes our theory of c-mode oscillations of accretion disks. Its quantitative accuracy is limited by the WKB approximation used throughout, as in previous papers \citep{nw92,per}. On the other hand, its qualitative results, that is, the above properties of the c-mode spectrum and its dependence on the rotation parameter $a$, rest on a single assumption that the speed of sound does not drop too fast with radius, which seems to be valid for existing models of accretion disks. If, nevertheless, such a fast drop does take place [$\nu>5/4$ in our terms, see formula (\ref{eq:33})], then the qualitative picture changes drastically. The set of the radial mode numbers becomes infinite, there are no cutoff values of the rotation parameter $a$, and the c-modes of the studied type are present in any corotating disk, no matter how slow its rotation. As it turns out, for $\nu<5/4$ the same effect on the results is caused by using a very rough approximation of the radial eigenfunction, namely, just the Airy function (\ref{eq:35}) throughout the whole trapping zone \citep{sw}: in this approximation the number of the radial modes becomes infinite, and the cutoff values of parameter $a$ disappear, so that all the modes exist for any corotating disk.

\section{Numerical Results and Discussion}

We now apply the above analysis to models of black hole accretion disks. The numerical results discussed below were obtained by assuming that gas pressure dominates within the disk, so that $\Gamma = 5/3$. However, this is only true near the inner edge $r_i$ [justifying our choice $\mu=2/5$ in equation (\ref{eq:33}) consistent with the results of \citet{pt}], at radii $r\gg r_i$, and at all radii for luminosities $L\ll L_{Edd}$. 

Recall that the interior structure of the (zero buoyancy) accretion disk enters our formulation only through the constant $\Gamma$ and the function $\alpha(r)$ (not to be confused with the usual viscosity parameter, here denoted by $\alpha_*$), inversely proportional to the speed of sound on the midplane and modeled by equation (\ref{eq:33}). In this equation we also chose $\gamma(r)$ to be constant, and took $\nu=0$ unless otherwise indicated. In section 4.3, we found that the properties of the modes do not change greatly when small amounts of buoyancy are introduced.

For all of the tables (but not the figures), we used $M=10^8M_\sun$, $\alpha_*=0.01$, and $L=0.1L_{Edd}$ so we could compare our results with those of \citet{p}. For a standard thin accretion disk \citep{ss}, at large radii (where the disk is nonrelativistic and gas pressure dominates) the speed of sound is given by
\be
c_s/c = 2.21\times 10^{-2}\frac{(L/L_{Edd})^{1/5}}{(\alpha_*M/M_\sun)^{1/10}} (rc^2/GM)^{-9/20} \qquad (r\gg r_i) \; , \label{eq:75}
\ee
corresponding to $\nu=9/20$ \citep{n}. (We are here reverting to ordinary units.) Since the physical conditions within the disk are more uncertain near its inner edge (but where gas pressure should also dominate), we have also used the nonrelativistic expression  
\be
c_s/c = 4.32\times 10^{-3}\frac{(L/L_{Edd})^{1/5}}{(\alpha_*M/M_\sun)^{1/10}}[c^2(r-r_i)/GM]^{2/5} \qquad (r\cong r_i)  \label{eq:76}
\ee
there \citep{n}, as indicated previously. We employ this relation (\ref{eq:76}) when using $\nu=0$ in equation (\ref{eq:33}). We obtain the (constant) value of $\gamma$ used in equation (\ref{eq:33}) by setting $\alpha(r)=1/c_s(r)$ there (again neglecting the relativistic corrections near the inner edge). The various relativistic corrections only become significant when $a\gtrsim 0.5$.
We have only studied the fundamental c--mode ($m^2=j=1$), but have employed various values of the radial mode number $n$. 

In Table 1 we present the cutoff values of $a$ [$a_n$, from equation (\ref{eq:63}); for $a<a_n$, the mode does not exist] for radial mode numbers $n=0,1,2$ and two choices of the boundary condition parameter. There is no c--mode for $n=0$ and $\theta=0$; no solution is also indicated by no entry in the subsequent tables. The results for $\theta=\pi/4$ are essentially identical to those shown for $\theta=\pi/2$. The results agree with the analytic expressions (\ref{eq:64}) valid for values of $a_n\ll 1$. 
These results for $\theta=\pi/2$ are extended to larger values of $n$ in Table 2. The maximum $n=175$ included corresponds to the requirement $a<a_\perp\cong 0.953$ for the validity of our analysis.  

Table 3 presents values of our basic expansion parameter $\epsilon(r_i,r_c)$, obtained from equation (\ref{eq:20}), as a function of $a$ for the lower radial mode numbers and two choices of boundary condition. We see that it is indeed much less than unity, as also required for our analysis. Recall that $\epsilon(r_i,r_c) > \epsilon(r,r_c) > 0$ for all $r_c>r>r_i$. 
The same format is used for Tables 4 and 5. The eigenfrequency $\sigma$ (in units of $c^3/GM$, as usual) is presented in Table 4. We see that it is a decreasing function of the radial mode number $n$, more strongly for lower values of $a$. The corresponding radial extent $r_c-r_i$ of the mode (in units of $GM/c^2$, as usual) is shown in Table 5. It increases significantly with radial mode number. We note that the sensitivity of the results in Tables 3--5 to the boundary condition parameter $\theta$ is modest.

Table 6 shows the dependence of the outer radius $r_c$ of the capture zone on values of $a$ near its cutoff $a_n$, for $n=0$ and $\theta=\pi/2$ (corresponding to $a_0=3.22\times 10^{-5}$). From these data, one can see that when $\log(a-a_0)\lesssim -6$, the asymptotic formula (\ref{eq:69}) holds: 
$r_c\propto(a-a_0)^{-q}$. The exponent $q=0.56$ is very close to that predicted, $q=4/7$, for $\nu=0$.  

The effect of changing the behavior of the speed of sound at large radii is illustrated in Table 7. The radial size of the mode for $\nu=0$ (from Table 5) is compared to that obtained with the choice $\nu=9/20$ of the standard accretion disk model, equation (\ref{eq:75}). This shows that the decrease in the speed of sound with radius also reduces the size of the trapping region, as expected. Correspondingly, the values of $a_n$ shown in Table 1 are reduced by about a factor of 7.5, and the values of $\epsilon(r_i,r_c)$ in Table 3 are reduced by about a factor of 2. 
The eigenfrequencies in Table 4 are increased by factors of 1.2--2.8 (with $n=0,1$) for $a=10^{-3}$, but were unaffected for $a=0.5$, as expected.

We present our major observationally relevant results in the following figures. For them, we have chosen the radial mode number $n=0$ and the boundary condition parameter $\theta=\pi/2$. (One might expect that the lowest radial mode could be the one most easily excited and with the largest net modulation.) The speed of sound parameter is taken to be $\nu=0$ for Figures 2 and 3, and $\nu=9/20$ for Figures 4(a,b) (which differed very little when $\nu=0$ was used). 

In Figure 2, we plot the relation between the fundamental c--mode and g--mode eigenfrequencies (scaled by the black hole mass) and the black hole angular momentum. This illustrates the dramatically different dependence of the c--mode frequency. Also shown is the orbital frequency $\Omega_{max}=\Omega(r_i)$ of a (commonly invoked) `blob' at the inner disk radius.  As expected, the c--mode frequency approaches $\Omega_{max}/2\pi$ as $a\rightarrow 1$. The case $\nu=9/20$ lies almost entirely within the band containing the range of masses and luminosities indicated. It is significant that the c--mode results of \citet{p}, obtained by numerical integration of equations (\ref{eq:4}) and (\ref{eq:5}), agree within the band shown in this figure (obtained via a further radial WKB approximation).

In Figure 3, we present our numerical results (points) for the size of the trapping region. The curves shown are the fits of these results to the formula
\be
r_c-r_i = K_0(GM/c^2)a^{-K_1}(1-a)^{K_2} \; , \label{eq:77}
\ee
giving
\begin{eqnarray*}
K_0 & = & 0.058\; , \quad K_1=0.66\; , \quad K_2=0.31\; \quad (M=10M_\sun)\; , \\
K_0 & = & 0.021\; , \quad K_1=0.55\; , \quad K_2=0.39\; \quad (M=10^8M_\sun)\; .
\end{eqnarray*}
This illustrates the fact that the radial extent of the mode is only appreciable for slowly rotating black holes ($a\ll 1$). To obtain the corresponding eigenfrequency, one can then use equation (\ref{eq:77}) and the known function $r_i=Mf(a)$ to obtain the value of the trapping radius $r_c$. Our fundamental result, equation (\ref{eq:17}) [or equation (\ref{eq:19}) for $a\ll 1$] with $m=-1$, then gives $|\sigma|$.

The c--mode frequency (scaled by mass) depends only on the black hole angular momentum and the accretion disk speed of sound. The first dependence has been shown in Figure 2. Rather than showing the second dependence directly, it is more relevant to instead use an observable, the luminosity. For fixed $\alpha_*$ and $M$, we use equation (\ref{eq:75}) to relate it to the speed of sound. We want to know how the frequency changes as the luminosity (proportional to the mass accretion rate) varies with time. That dependence is known to be very weak for the g--modes \citep{per}. In contrast, the $m=0$ fundamental p--mode has a frequency $|\sigma|\propto c_s^{1/3}$ and a radial size $(r_- - r_i) \propto c_s^{2/3}$ \citep{kf,nw91,p}.

We fit the dependences of the eigenfrequency on accretion disk luminosity shown by the points in Figures 4(a) and 4(b) by the form $|\sigma|\propto L^{-K(M,a)}$. The results shown correspond to 
\begin{eqnarray*}
K(10M_\sun,10^{-3}) & = & 0.30\phn\; , \quad K(10^8M_\sun,10^{-3}) = 0.039\phn \; ;\\ K(10M_\sun,10^{-1}) & = & 0.020 \; , \quad K(10^8M_\sun,10^{-1}) = 0.0044 \; .
\end{eqnarray*}
Note that the dependence is weaker for the larger value of $a$, again as expected since the properties of the disk (except its inner radius) become irrelevant as $a\rightarrow 1$.

Although the fundamental c-mode is almost incompressible, the changing projected area of the mode could modulate the luminosity via reflection of radiation from the postulated `corona' surrounding the accretion disk. Of course, this requires that the disk not be viewed close to face-on. The observability of a c--mode induced modulation of the detected flux would seem to require large values of $r_c-r_i$. This in turn would imply small values of $a$, from Figure 3, and correspondingly small values of frequency, from Figure 2. We then see from Figure 4(a) that the dependence of the frequency on luminosity is relatively weak but might be detectable for the stellar mass black holes. Issues such as the excitation and damping of the c--modes, including their leakage into the black hole via accretion from the inner edge of the  disk, are beyond the scope of this paper.

Finally, we emphasize that of the fundamental (g, p, c) modes, only the c--mode can (generically) have a frequency $|\sigma|\ll\Omega$. The c--modes are candidates for those low frequency features in the power spectra of accreting black holes whose frequency varies only weakly with changes in luminosity. An example of a candidate is the 9 Hz modulation in the `microquasar' GRO J1655-40 observed by the RXTE satellite \citep{rem}. Since the mass of the presumed black hole in this binary has been determined to be $M\cong 7M_\sun$ \citep{sha}, this frequency requires a black hole angular momentum $a\cong 0.18$ (Figure 2) if it is produced by a low $n$ c--mode. 

However, we see from Figure 3 that the lowest radial mode would have a radial extent of only $0.2GM/c^2$. (But from the results in Table 5, we expect the radial extent of the $n=1$ mode to be about 3 times greater.) In addition, the energy spectrum of the X-rays when this modulation was present was `softer' (more thermal) than typical. This may complicate the above proposal that the photons are modulated via disk reflection from the corona. Clearly, one should search for other radial (or vertical) modes to confirm any identification.

\acknowledgments

This work was supported by NASA grant NAS 8-39225 to Gravity Probe B and NASA grant NAG 5-3102 to RVW, who also thanks the Aspen Center for Physics for support during a 1999 summer workshop. We are grateful to Lev Kapitanski for finding the number theory results for the Diophantine equation (\ref{eq:12}), and to Dana Lehr for her remarks and help with the figures.

\begin{deluxetable}{c c c c}
\tablewidth{0pt}
\tablecaption{Cutoff values ($a_n$) of $a$}
\tablehead{\colhead{$\theta$} &
           \colhead{$a_0$}    &
           \colhead{$a_1$}    &
           \colhead{$a_2$}    }
\startdata
$\pi/2$  &  $3.2 \ 10^{-5}$ & $3.7 \ 10^{-4}$ & $1.0 \ 10^{-3}$ \\
$0$      &   \nodata        & $6.3 \ 10^{-5}$ & $4.6 \ 10^{-4}$ 
\enddata
\end{deluxetable}

\begin{deluxetable}{c c c c c c c c}
\tablecolumns{8}
\tablewidth{0pc}
\tablecaption{Cutoff values of $a$, for $\theta=\pi/2$}
\tablehead{
\colhead{$n$} &
\colhead{$a_n$} &
\colhead{ } &
\colhead{$n$} &
\colhead{$a_n$} &
\colhead{ } &
\colhead{$n$} &
\colhead{$a_n$} }

\startdata
$  0$  & $     3.25 \ 10^{-5} $ & & 
$  7$   & $    1.05 \ 10^{-2}$ & &
$40 $     & $ 2.35 \ 10^{-1} $ \\

$  1$   &  $   3.75 \ 10^{-4}$ & &
$  8$    & $   1.25 \ 10^{-2}$ & &
$60 $ & $     4.15 \ 10^{-1}$ \\

$  2$    &  $  1.05 \ 10^{-3}$ & &
$  9$     & $  1.65 \ 10^{-2}$ & &
$75 $  & $    5.35 \ 10^{-1}$ \\

$  3$     &  $ 2.15 \ 10^{-3}$ & &
$10 $ & $     1.95 \ 10^{-2} $ & &
$100$   & $   6.95 \  10^{-1}$ \\

$  4$      & $ 3.65 \ 10^{-3}$ & &
$11 $  & $    2.35 \ 10^{-2} $ & &
$125$    & $  8.25 \ 10^{-1}$ \\

$  5$ & $      5.45 \ 10^{-3}$ & &
$12 $   & $   2.75 \ 10^{-2} $ & &
$150$     & $ 8.95 \ 10^{-1}$ \\

$  6$  & $     7.55 \ 10^{-3}$ & &
$20 $    & $  7.15 \ 10^{-2} $ & &
$175$     & $ 9.45 \ 10^{-1}$ 
\enddata

\end{deluxetable}

\begin{deluxetable}{ccccccccc} 
\tablecolumns{8} 
\tablewidth{0pc} 
\tablecaption{Expansion parameter $\epsilon(r_i,r_c)$} 
\tablehead{ 
\colhead{} &
\colhead{} & 
\colhead{} & 
\multicolumn{6}{c}{$a$} \\ 
\cline{4-9} \\ 
\colhead{$n$} & \colhead{$\theta$} & 
\colhead{ } &
\colhead{$10^{-4}$}    & 
\colhead{$10^{-3}$}    & 
\colhead{$10^{-2}$}    & 
\colhead{$10^{-1}$}    & 
\colhead{$0.5$}    & 
\colhead{$0.95$}  }     
\startdata 
0 & $\pi/2$ & & $1.0 \ 10^{-5}$ & $4.4 \ 10^{-5}$ & $1.6 \ 10^{-4}$ & $5.9 \ 10^{-4}$     & $1.7 \ 10^{-3}$  & $6.4 \ 10^{-3}$ \\
0 & $0$     & & \nd           & \nd & \nd      & \nd       & \nd       & \nd \\
1 & $\pi/2$ & & \nd           & $1.1 \ 10^{-4}$ & $4.7 \ 10^{-4}$ & $1.8 \ 10^{-3}  $   & $5.3 \ 10^{-3} $    & $1.9 \ 10^{-2}$\\
1 & $0$     & & $1.3 \ 10^{-5}$ & $5.7 \ 10^{-5}$ &$ 2.2 \ 10^{-4}$ &$ 7.5 \ 10^{-4}$     & $2.4 \ 10^{-3}  $   & $8.7 \ 10^{-3}$ \\
2 & $\pi/2$ & & \nd           & \nd           & $7.0 \ 10^{-4}$ & $2.8 \ 10^{-3}  $   & $8.5 \ 10^{-3}$ & $3.1 \ 10^{-2}$ \\
2 & $0$     & & \nd           & $1.2 \ 10^{-4}$ & $5.1 \ 10^{-4}$  & $2.0 \ 10^{-3}$    & $5.8 \ 10^{-3} $    & $2.1 \ 10^{-2}$ 
\enddata 
\end{deluxetable} 

\begin{deluxetable}{ccccccccc} 
\tablecolumns{8} 
\tablewidth{0pc} 
\tablecaption{Eigenfrequencies $-\sigma/(c^3/GM)$} 
\tablehead{ 
\colhead{} &
\colhead{} & 
\colhead{} & 
\multicolumn{6}{c}{$a$} \\ 
\cline{4-9} \\ 
\colhead{$n$} & \colhead{$\theta$} & 
\colhead{ } &
\colhead{$10^{-4}$}    & 
\colhead{$10^{-3}$}    & 
\colhead{$10^{-2}$}    & 
\colhead{$10^{-1}$}    & 
\colhead{$0.5$}    & 
\colhead{$0.95$}  }     
\startdata 
0 & $\pi/2$ & & $2.1 \ 10^{-7}$ & $6.3 \ 10^{-6}$ & $8.3 \ 10^{-5}$ & $1.0 \ 10^{-3}$     & $0.011\phn$  & $0.12$ \\
0 & $0$     & & \nd           & \nd & \nd      & \nd       & \nd       & \nd \\
1 & $\pi/2$ & & \nd           & $1.8 \ 10^{-6}$ & $6.2 \ 10^{-5}$ & $9.3 \ 10^{-4}  $   & $0.010\phn $    & $0.12$\\
1 & $0$     & & $6.0 \ 10^{-8}$ & $5.4 \ 10^{-6}$ &$ 7.9 \ 10^{-5}$ &$ 1.0 \ 10^{-3}$     & $0.010\phn$     & $0.12$ \\
2 & $\pi/2$ & & \nd           & \nd           & $4.6 \ 10^{-5}$ & $8.6 \ 10^{-4}  $   & $0.0099$ & $0.12$ \\
2 & $0$     & & \nd           & $1.2 \ 10^{-6}$ & $5.9 \ 10^{-5}$  & $9.2 \ 10^{-4}$    & $0.010\phn $    & $0.12$ 
\enddata 
\end{deluxetable} 

\begin{deluxetable}{ccccccccc} 
\tablecolumns{8} 
\tablewidth{0pc} 
\tablecaption{Trapping region $(r_c-r_i)/(GM/c^2)$} 
\tablehead{ 
\colhead{} &
\colhead{} & 
\colhead{} & 
\multicolumn{6}{c}{$a$} \\ 
\cline{4-9} \\ 
\colhead{$n$} & \colhead{$\theta$} & 
\colhead{ } &
\colhead{$10^{-4}$}    & 
\colhead{$10^{-3}$}    & 
\colhead{$10^{-2}$}    & 
\colhead{$10^{-1}$}    & 
\colhead{$0.5$}    & 
\colhead{$0.95$}  }     
\startdata 
0 & $\pi/2$ & & 3.8 & 0.83     & 0.25     & 0.079     & 0.023     & 0.0064 \\
0 & $0$     & & \nd & \nd      & \nd      & \nd       & \nd       & \nd \\
1 & $\pi/2$ & & \nd & 4.4\phn  & 0.89     & 0.25\phn  & 0.073     & 0.019\phn\\
1 & $0$     & & 8.9 & 1.2\phn  & 0.35     & 0.10\phn  & 0.032     & 0.0087 \\
2 & $\pi/2$ & & \nd & \nd      & 1.6\phn  & 0.42\phn  & 0.12\phn  & 0.032\phn\\
2 & $0$     & & \nd & 5.7\phn  & 1.0\phn  & 0.28\phn  & 0.081     & 0.021\phn 
\enddata 
\end{deluxetable} 

\begin{deluxetable}{c c c c c}

\tablewidth{0pt}

\tablecaption{Capture radius for $a$ near its lower limit ($n=0$)}

\tablehead{
\colhead{$\log (a - a_0)$} &
\colhead{$\log r_c$} &
\colhead{ }  &           
\colhead{$\log (a - a_0)$} &
\colhead{$\log r_c$} }
      
\startdata
-5\phd \phn \phn & 1.29 & & -6.75 & 2.22 \\
-5.25 & 1.41 & & -7$\phantom{.00}$   & 2.36 \\
-5.5\phn  & 1.54 & & -7.25 & 2.50 \\
-5.75 & 1.67 & & -7.5\phn  & 2.64 \\
-6$\phantom{.00}$   & 1.80 & & -7.75 & 2.78 \\
-6.25 & 1.94 & & -8$\phantom{.00}$   & 2.92 \\
-6.5\phn  & 2.08 & & -8.25 & 3.06
\enddata

\end{deluxetable}

\begin{deluxetable}{cccccc} 
\tablecolumns{6} 
\tablewidth{0pc} 
\tablecaption{Dependence of $(r_c-r_i)/(GM/c^2)$ on $\nu$} 
\tablehead{ 
\colhead{} &
\colhead{} & 
\colhead{} & 
\colhead{} &
\multicolumn{2}{c}{$a$} \\ 
\cline{5-6} \\ 
\colhead{$\nu$} &
\colhead{$n$} & \colhead{$\theta$} & 
\colhead{ } &
\colhead{$10^{-3}$}    & 
\colhead{$0.5$}    }     
\startdata 
0&0 & $\pi/2$ & & $0.83$ & $0.023$  \\
0&0 & $0$     & & \nd    & \nd \\
0&1 & $\pi/2$ & & $4.4\phn$ & $0.073$  \\
0&1 & $0$     & & $1.2\phn$ & $0.032$  \\
9/20&0 & $\pi/2$ & & $0.35$ & $0.013$  \\
9/20&0 & $0$     & & \nd    & \nd \\
9/20 &1& $\pi/2$ & & $1.3\phn$ & $0.040$  \\
9/20&1 & $0$     & & $0.50$ & $0.017$ 
\enddata 
\end{deluxetable} 

\begin{figure}
\figurenum{1}
\epsscale{0.82}
\plotone{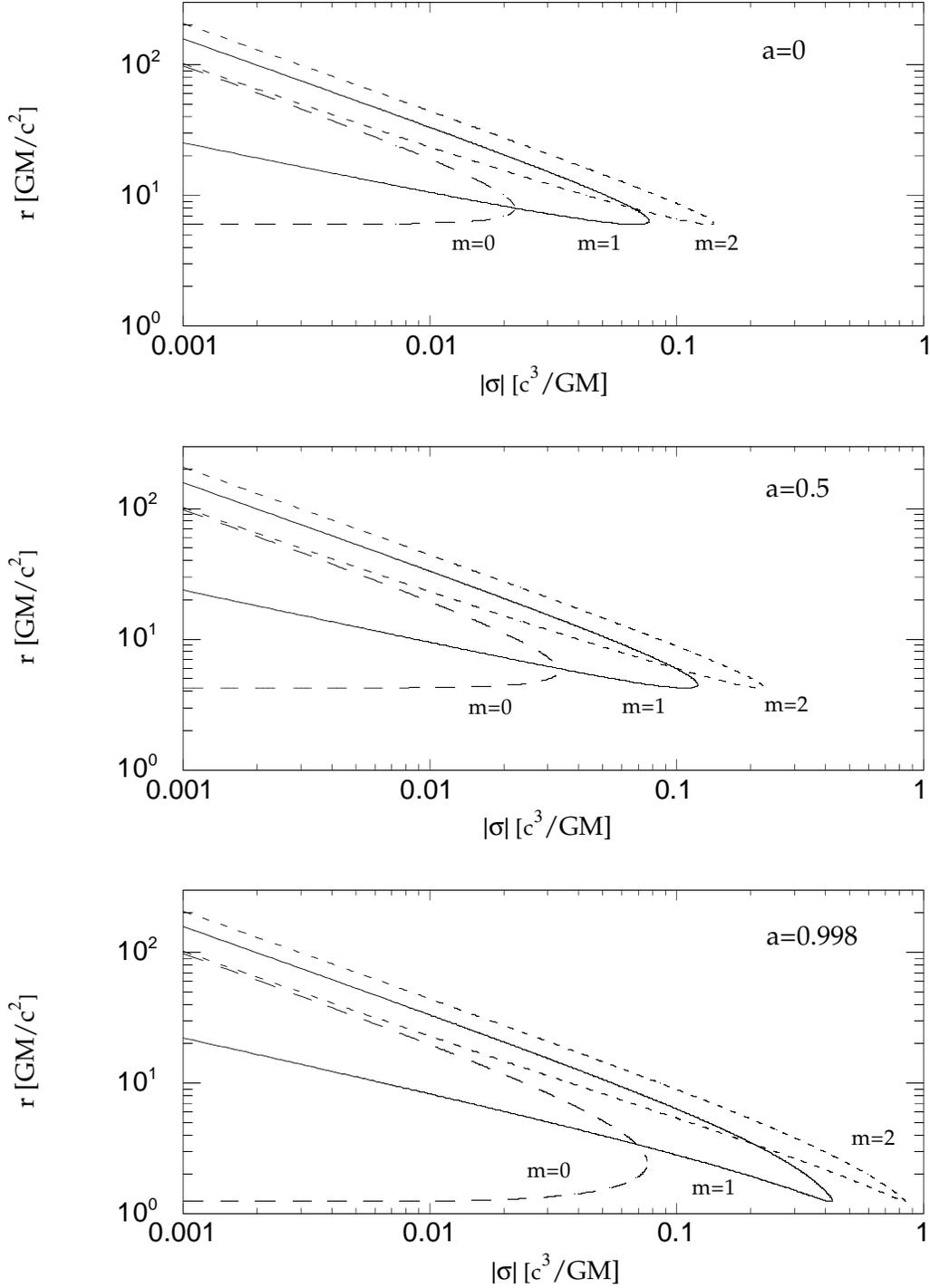}
\caption{Radii between which $\omega^2<\kappa^2$, as a function of the eigenfrequency. Three choices of axial mode number $m$ and black hole (dimensionless) angular momentum $a$ are included.}
\end{figure}

\begin{figure}
\figurenum{2}
\epsscale{0.98}
\plotone{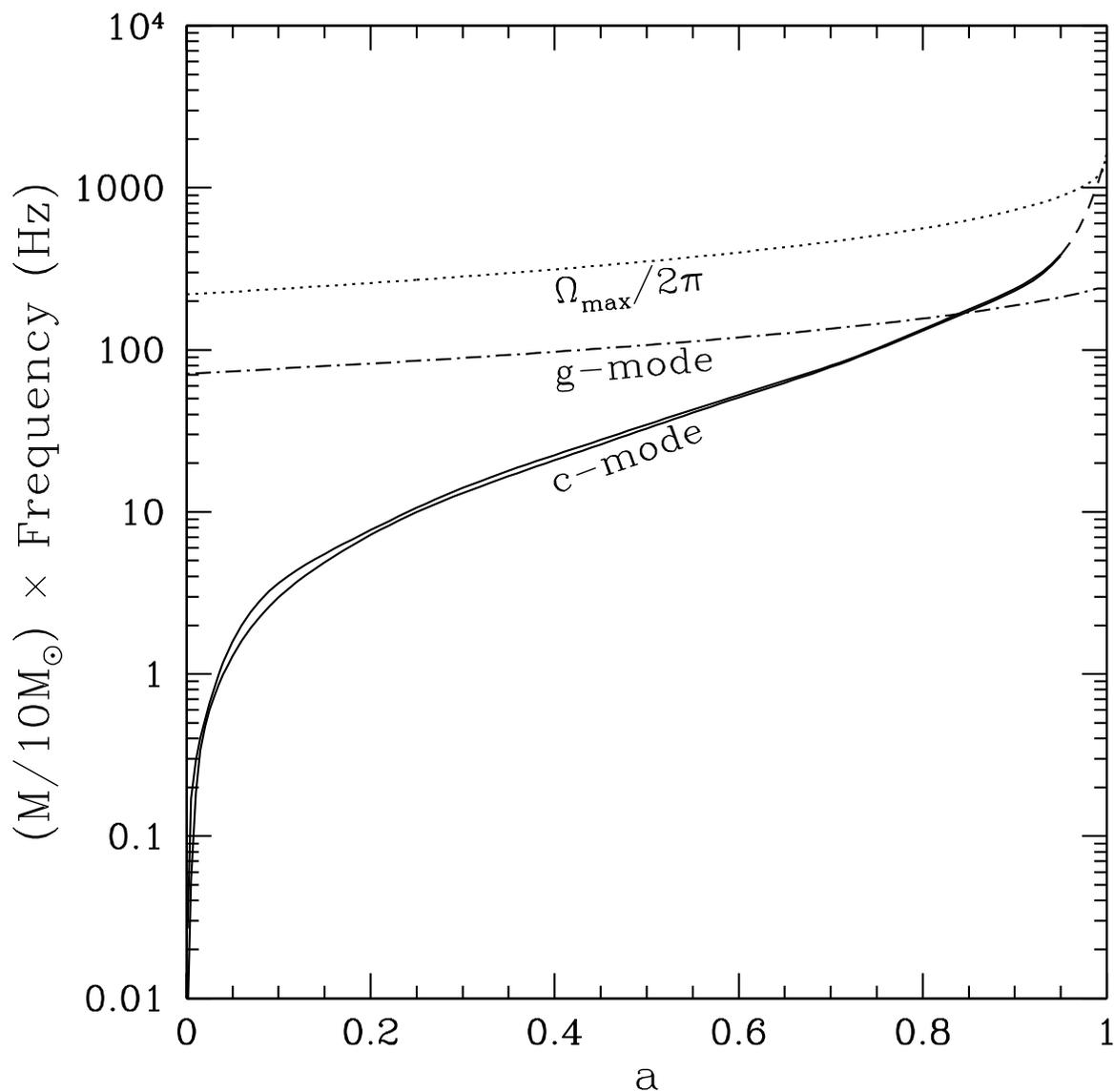}
\caption{The dependence of three characteristic disk frequencies on the angular momentum of the black hole. The top one is the orbital frequency of a `blob' at the inner edge of the accretion disk. The c--mode curves bound the region corresponding to the parameter ranges $0.001\leq L/L_{Edd}\leq 0.1$ and $10\leq M/M_\sun\leq 10^9$, with $\alpha_*=0.1$.}
\end{figure}

\begin{figure}
\figurenum{3}
\epsscale{0.98}
\plotone{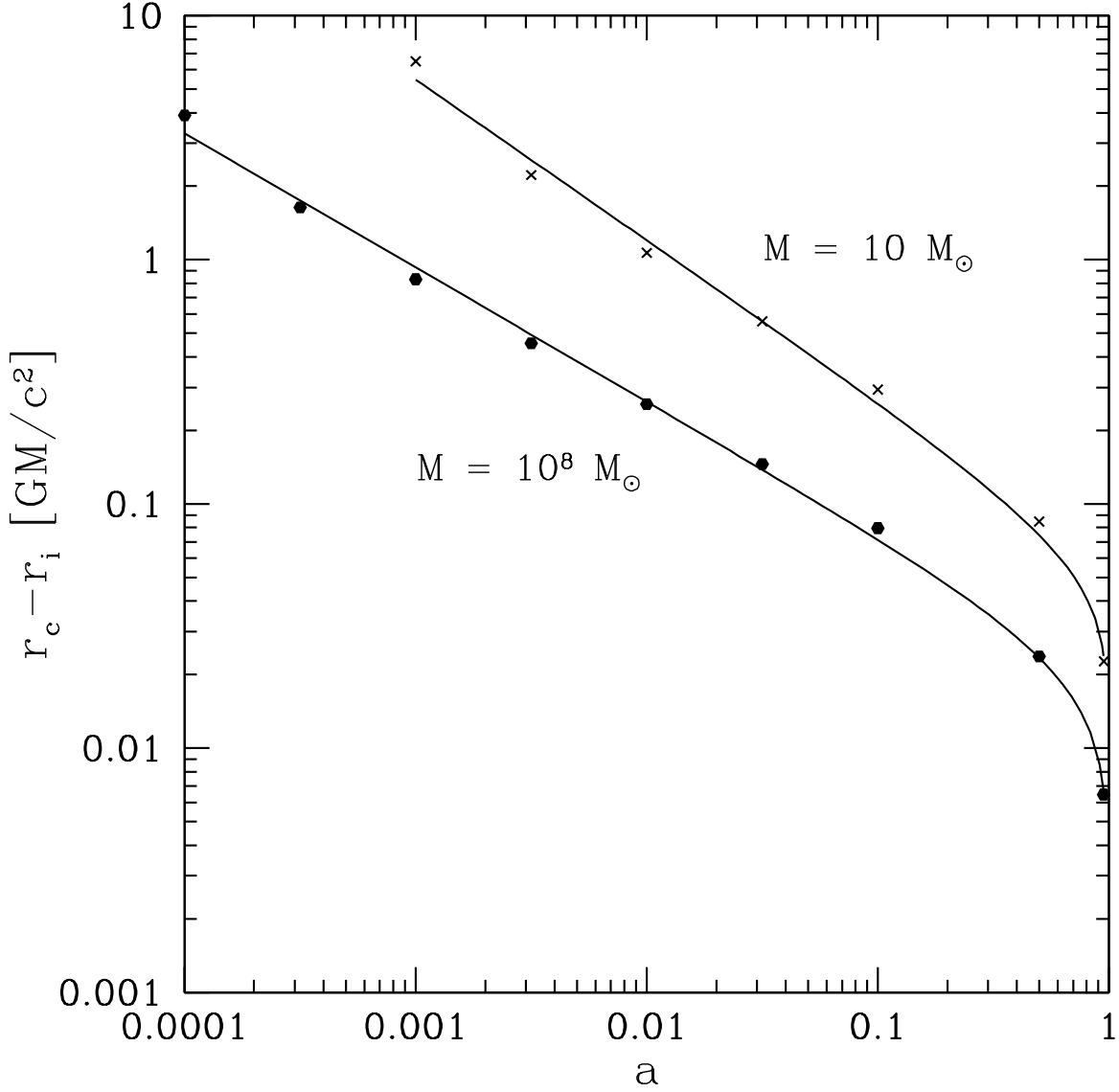}
\caption{The dependence of the radial extent of the mode trapping region on the angular momentum of the black hole, choosing $L=0.1 L_{Edd}$. For $M=10 M_\sun$ (crosses), $\alpha_*=0.1$, while for $M=10^8 M_\sun$ (hexagons), $\alpha_*=0.01$. The fits are of the form $K_0a^{-K_1}(1-a)^{K_2}$.}
\end{figure}

\begin{figure}
\figurenum{4}
\epsscale{1.10}
\plottwo{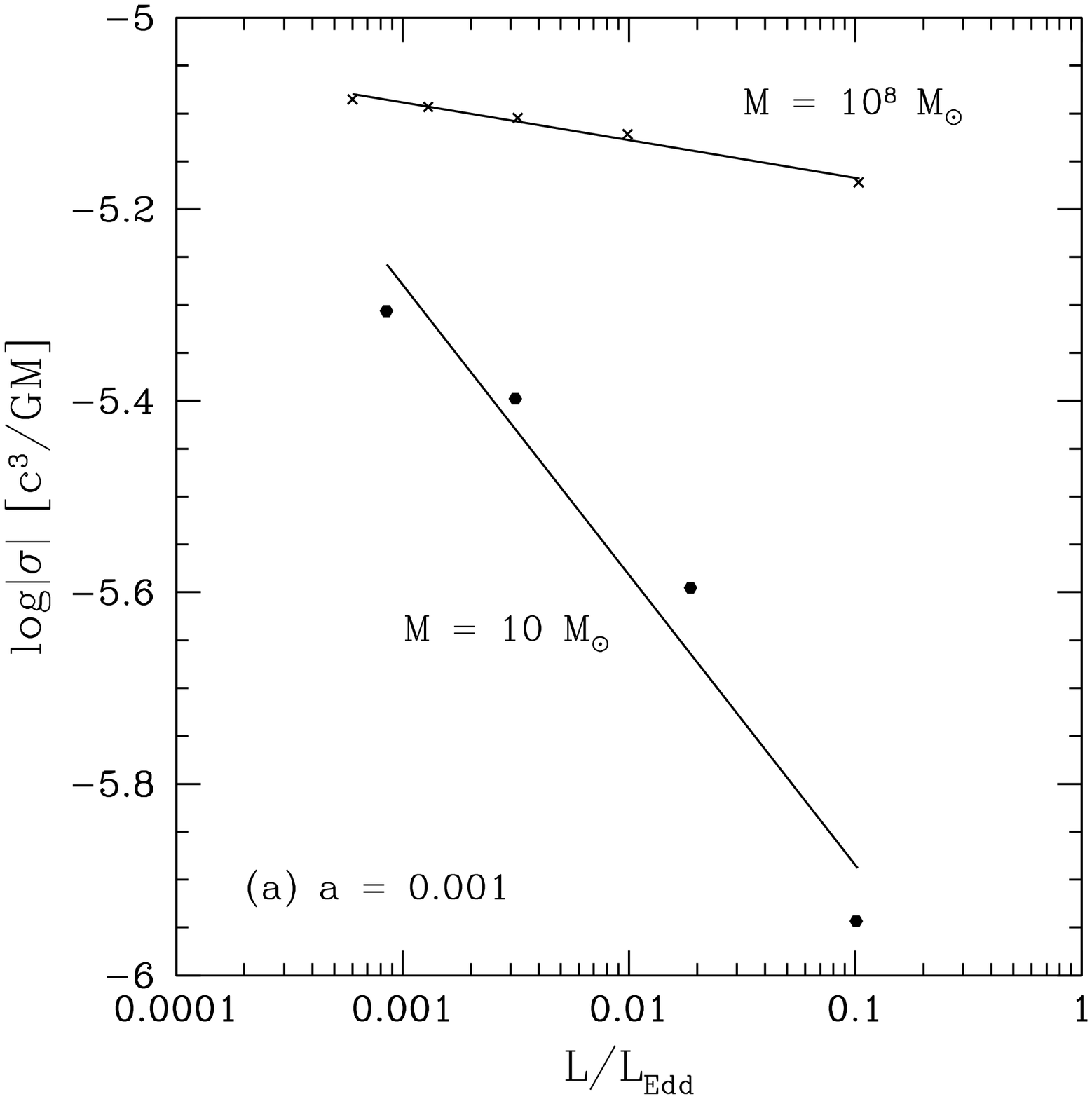}{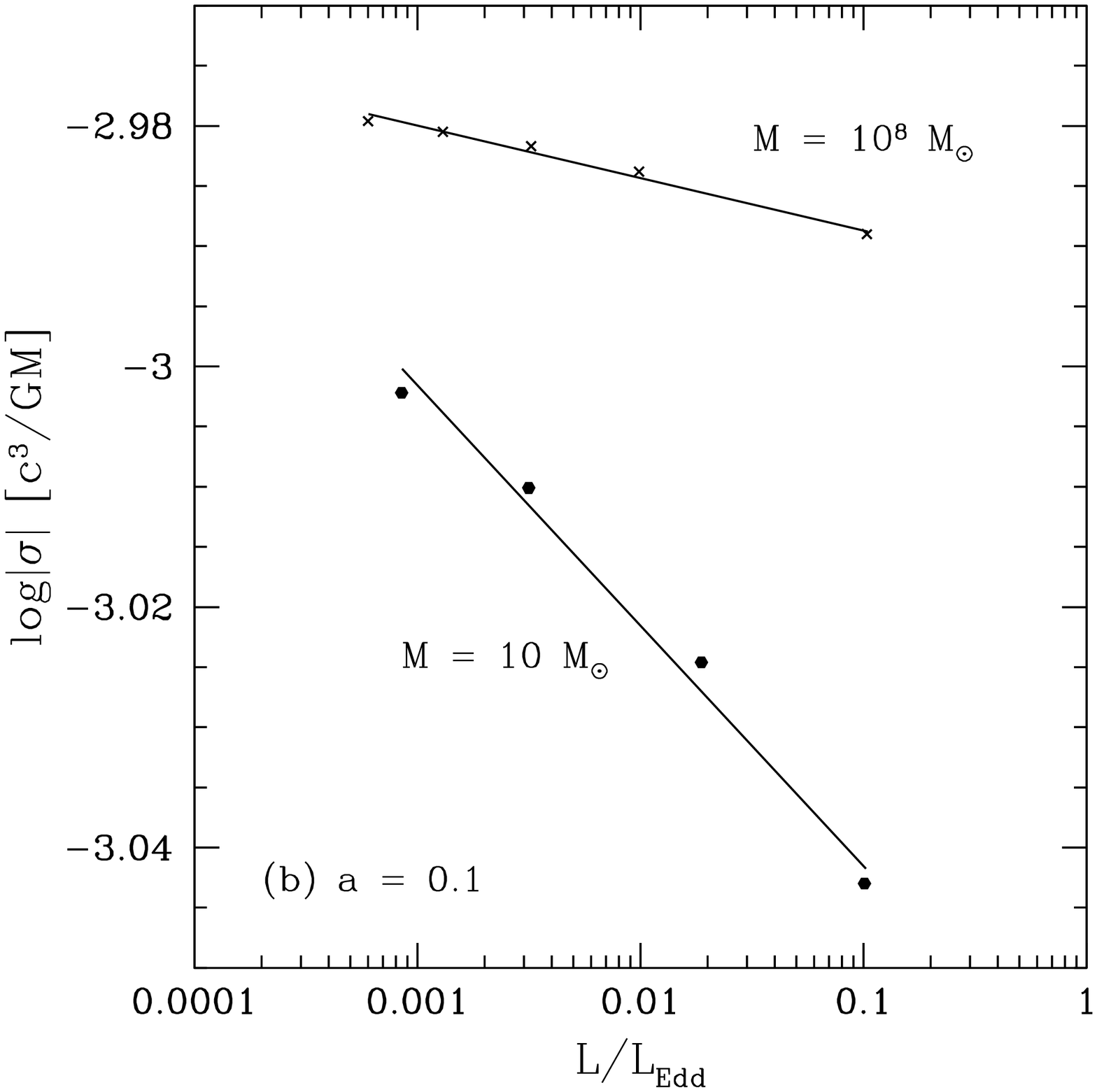}
\caption{The dependence of the eigenfrequency on the accretion disk luminosity, choosing $\alpha_*=0.1$. For plot (a), $a=0.001$ and for plot (b), $a=0.1$.}
\end{figure}

\end{document}